\documentclass[journal]{ieeetran}
\usepackage{cite}
\usepackage[pdftex]{graphicx}
\usepackage{epstopdf}
\usepackage{amsmath}
\usepackage{cases}
\usepackage[caption=false,font=footnotesize]{subfig}
\usepackage{fixltx2e}
\usepackage{amssymb}
\usepackage{mathtools}
\usepackage{bm}
\usepackage{multirow}
\usepackage{color}
\usepackage[ruled]{algorithm2e}
\usepackage{mathrsfs}
\usepackage{amsthm,amsmath,amssymb}
\usepackage{threeparttable}
\usepackage{url}
\usepackage{booktabs}
\usepackage{bbding}
\usepackage{wasysym}
\newcommand{\tabincell}[2]{\renewcommand\arraystretch{0.9}\begin{tabular}{@{}#1@{}}#2\end{tabular}}

\begin{document}

\title{Optimal Sizing of Isolated Renewable Power Systems with Ammonia Synthesis: Model and Solution Approach}

\author{Zhipeng~Yu,
        Jin~Lin,~\IEEEmembership{Member,~IEEE},
        Feng~Liu,~\IEEEmembership{Senior Member,~IEEE},
        Jiarong~Li,
        Yuxuan~Zhao, and
        Yonghua~Song,~\IEEEmembership{Fellow,~IEEE}

\thanks{Financial support came from the National Key R\&D Program of China (2021YFB4000500). \emph{(Corresponding author: Jin Lin)}
}
\thanks{Zhipeng Yu, Jin Lin, Feng Liu, Jiarong Li, and Yuxuan Zhao are with the State Key Laboratory of Control and Simulation of Power Systems and Generation Equipment, Department of Electrical Engineering, Tsinghua University, Beijing 100087, China. And Jin Lin is also with Sichuan Energy Internet Reasearch Institute, Tsinghua University, Chengdu, 610213, China. (e-mail: linjin@tsinghua.edu.cn) 
}
\thanks{Yonghua Song is with the Department of Electrical and Computer Engineering, University of Macau, Macau 999078, China, and also with the Department of Electrical Engineering, Tsinghua University, Beijing 100087, China. 
}
}

\maketitle

\begin{abstract}
   Isolated renewable power to ammonia (IRePtA) has been recognized as a promising way to decarbonize the chemical industry. Optimal sizing of the renewable power system is significant to improve the techno-economic of IRePtA since the investment of power sources exceeds 80\% of the total investment. However, multi-timescale electricity, hydrogen, and ammonia storages, minimum power supply for system safety, and the multi-year uncertainty of renewable generation lead to difficulties in planning. To address the issues above, an IGDT-MILFP model is proposed. First, the levelized cost of ammonia (LCOA) is directly formulated as the objective, rendering a mixed integer linear fractional programming (MILFP) problem. Information gap decision theory (IGDT) is utilized to handle the multi-year uncertainty of renewable generation. Second, a combined Charnes-Cooper (C\&C) transformation and Branch-and-Bound (B\&B) method is proposed to efficiently solve the large-scale IGDT-MILFP model, giving robust and opportunistic planning results. Then, Markov Chain Monte Carlo (MCMC) sampling-based posteriori analysis is leveraged to quantify the long-run performance. Finally, a real-life system in Inner Mongolia, China, is studied. The results indicate that the proposed methods could reduce the computational burden by orders of magnitude for solving a large-scale MILFP problem. Moreover, the proposed IGDT-MILFP model is necessary and accurate to obtain an optimal capacity allocation with the lowest expected LCOA (3610 RMB/t) in long-run simulations.
\end{abstract}


\section{Introduction}
\label{sec:intro}

\subsection{Background and Motivations}
\label{sec:background_motivation}
\begin{figure}[t]
   \centering
   \includegraphics[width=3.46in]{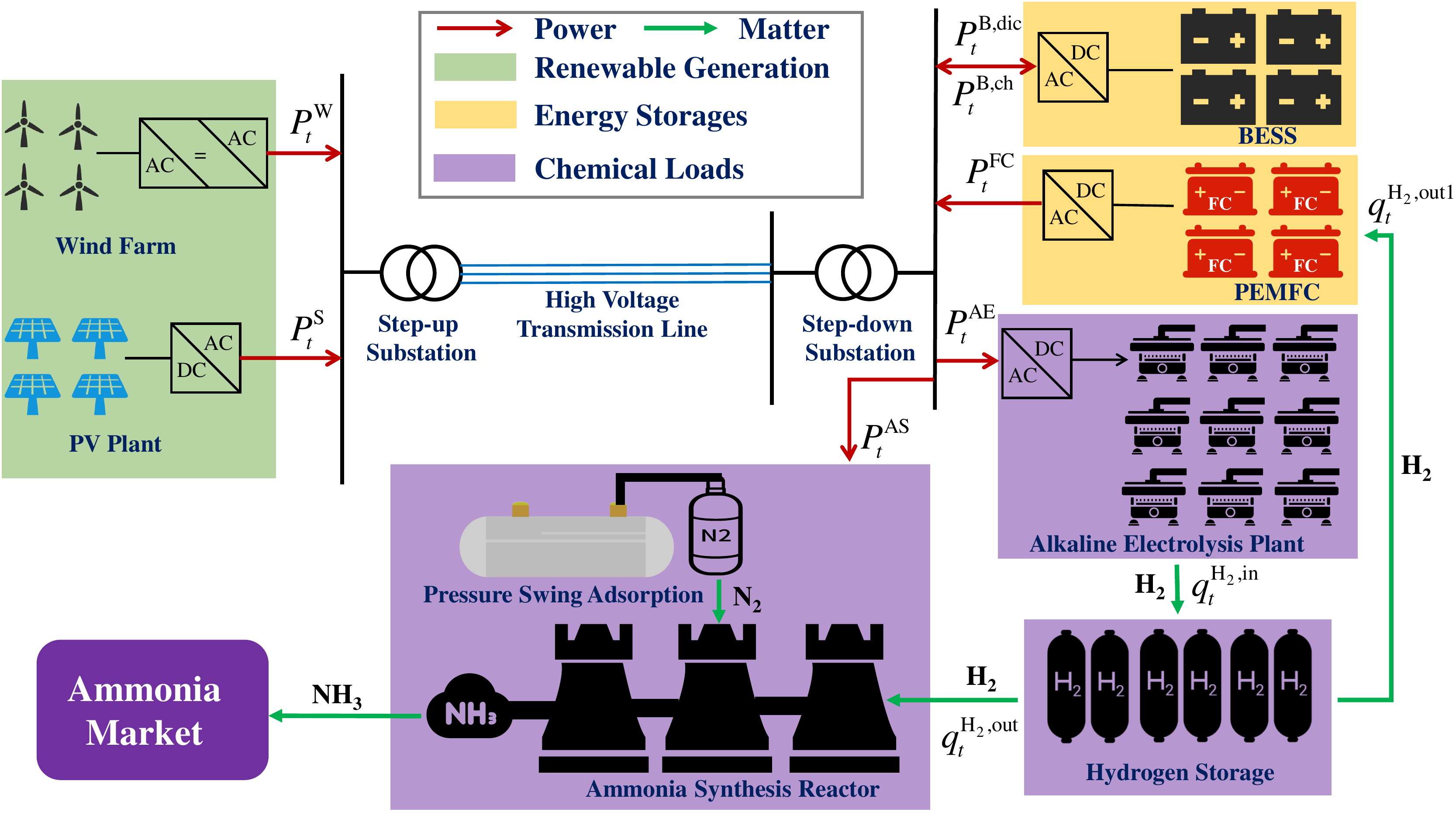}
   \caption{Isolated renewable power to ammonia system integration and operation mode.}
   \label{fig:Topu_IRePtA}
 \end{figure}

\IEEEPARstart{R}{enewable} power to ammonia (RePtA) is regarded as a promising way to decarbonize the chemical industry \cite{chehade2021progress,macfarlane2020roadmap}. The world's first green ammonia plant, at the commercial scale of 10 MW power, will be located in Western Jutland, Denmark, and produce more than 5000 t green ammonia per year from renewable power \cite{20215}. Such RePtA projects are multiplying in other countries, such as Saudi Arabia \cite{riera2022simulated}, Australia \cite{salmon2021impact}, and China \cite{Batou-Project-2021}. Some RePtA projects are planned to be supported by the power grid, known as grid-connected RePtA system \cite{Batou-Project-2021}. However, there are two problems, one is the difficulty of connecting to the power grid, and the other is the contradiction in the definition of a green fuel \cite{chiaramonti2019impacts} when too much electricity from the power grid is consumed. Hence isolated renewable power to ammonia (IRePtA), as shown in Fig. \ref{fig:Topu_IRePtA}, is a more prominent approach to realizing large-scale green ammonia production \cite{nayak2020techno,rouwenhorst2019islanded}. For example, China's Inner Mongolia has explicitly stated that off-grid is one of two allowed forms for RePtA planning since July 2022 \cite{Batou-Project-2022}.

In the IRePtA, the investment of the renewable power system (i.e., wind, solar, and energy storage) exceeds 80\% of the total investment, and the volatility of renewable generation significantly impacts system operation. Therefore, optimal sizing of the renewable power system plays a core role in improving the techno-economic of IRePtA. \cite{shepherd2022open} used a hybrid renewable energy source and battery energy storage system (BESS) as balancing technology and estimated a levelized cost of ammonia (LCOA) between 0.69 and 0.92 \$/kg $\rm{NH_3}$. However, in the study by \cite{shepherd2022open}, only electrolysis is modeled as a flexible load, but ammonia synthesis (AS) is regarded as a fixed load constantly operated at full load level, which significantly increases the demand for BESS and hydrogen storage and causes a large reduction in renewable energy. Additionally, the expensive investment cost of BESS causes LCOA to be much more costly than the market price ammonia. Therefore, exploiting the flexibility of power to ammonia (PtA), especially that of AS, is also essential to reduce the LCOA of the IRePtA system.

\begin{figure}[t]
   \centering
   \includegraphics[width=3.46in]{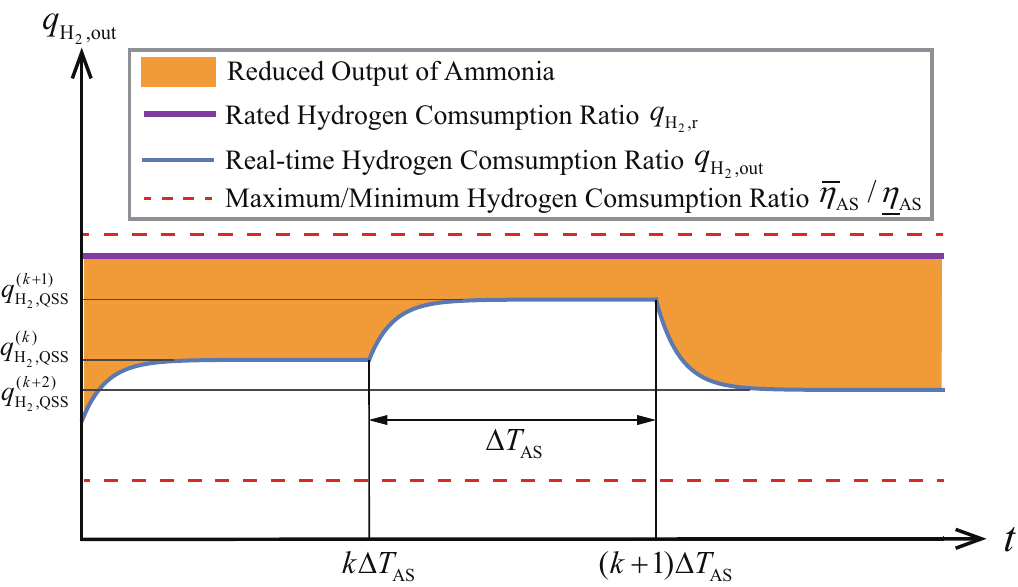}
   \caption{The proposed planned quasi-steady-state condition scheduling model for ammonia synthesis.}
   \label{fig:Ammonia_Adjustment_Fig}
 \end{figure}

Existing research works \cite{sanchez2018optimal,klyapovskiy2021optimal,armijo2020flexible} have demonstrated that AS can adjust the working condition in a wide range over a long duration. However, the dynamical regulation of AS will cause a reduction in the actual output of ammonia, as shown in the orange regions in Fig. \ref{fig:Ammonia_Adjustment_Fig}, and will in turn reduce the revenue for selling ammonia. How to plan an optimal capacity utilization rate of AS, to trade off flexibility and economy, is an emerging problem in practical engineering.

\begin{figure}[t]
    \centering
    \includegraphics[width=3.46in]{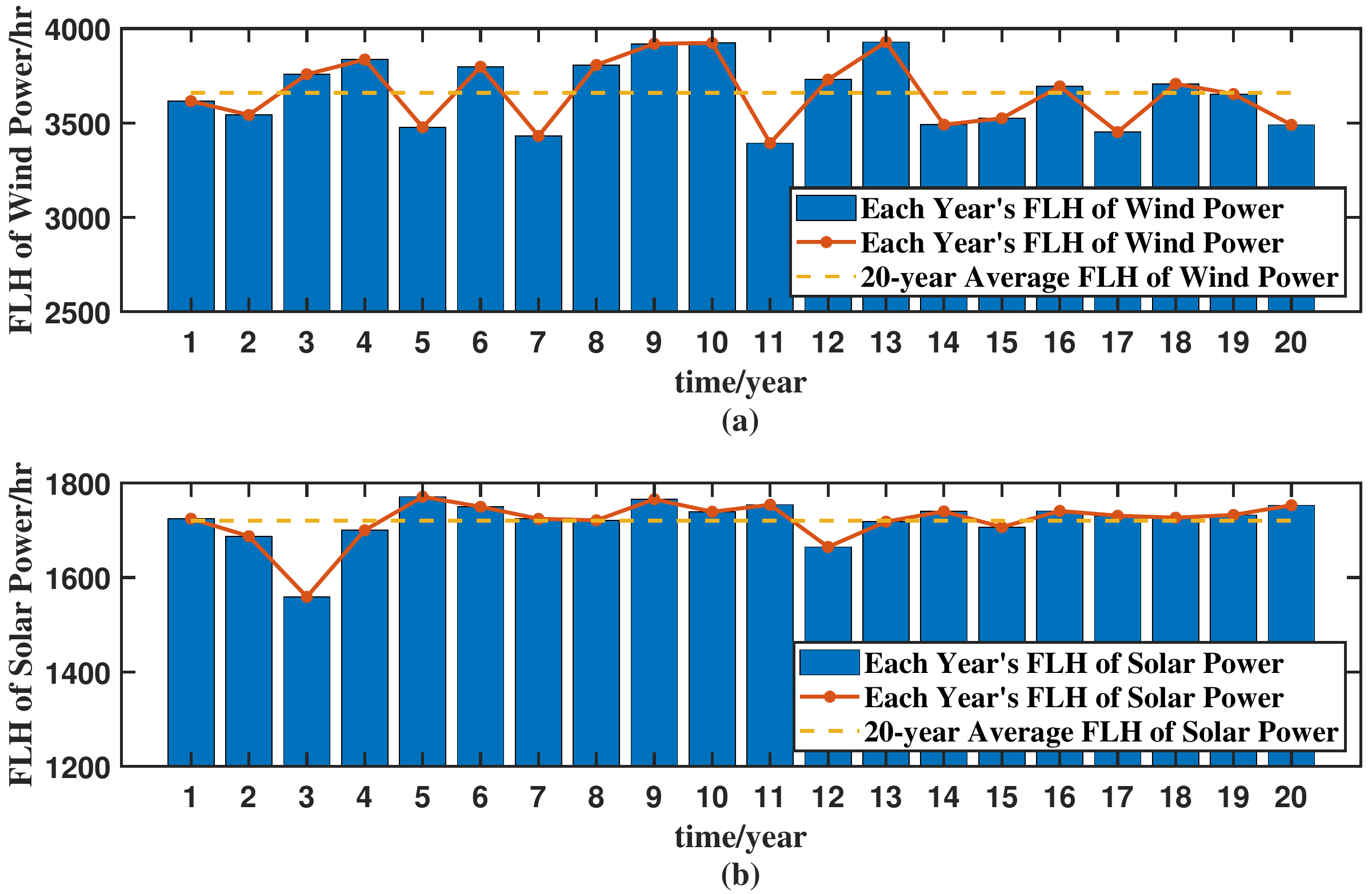}
    \caption{Multiple years volatility of renewable generation. (a) FLH of wind power; (b) FLH of solar power.}
    \label{fig:Multiyear_RG}
  \end{figure}

On the other hand, the planning of the IRePtA system is a typical long-run problem since the whole system's lifetime of operation is usually 20 years \cite{li2021co} or more. Therefore long-run risk derived from multi-year uncertainty of renewable generation, i.e., fluctuation of full load hours (FLH) of wind and solar power, as shown in Fig. \ref{fig:Multiyear_RG}, should be well considered. Addressing the above issues is the motivation of this paper.



\subsection{Literature Review}
\label{sec:literature_review}

Existing works \cite{pan2021cost,li2019optimal,li2021co,beerbuhl2015combined,verleysen2020can,li2022coordinated} have studied the capacity optimization of power to hydrogen (P2H) or PtA systems. \cite{pan2021cost} presented a detailed assessment of the levelized cost of hydrogen (LCOH) produced by a photovoltaic and grid-based hydrogen system (PGHS). Although minimizing LCOH is the objective, annual hydrogen production is given as a constant, evenly divided over 365 days, which is known as the deterministic hydrogen demand. \cite{li2019optimal} studied the optimal investment of electrolyzers and seasonal storage for hydrogen supply chains (HSC) incorporated with electric networks (EN) by minimizing the total investment and operation cost when predictive electricity demand and hydrogen demand are met. And \cite{li2021co} proposed a co-planning model for wind to ammonia (WtA) and EN by minimizing the total investment cost. Similarly, ammonia demand is given as a deterministic condition. \cite{beerbuhl2015combined} studied the capacity planning of electrolyzers and buffer tanks to maximize the net revenue.

These previous studies \cite{pan2021cost,li2019optimal,li2021co,beerbuhl2015combined} did not consider the impact of uncertainty on both capacity optimization and system operation. Furthermore, \cite{verleysen2020can} analyzed the robust design of a WtA system, considering the operational uncertainties, including wind speed measurement error and temperature variation. \cite{li2022coordinated} analyzed the coordinated planning of high-voltage direct current (HVDC) and power-to-hydrogen supply chains, fully considering renewable energy fluctuation (short-term uncertainty) and the price tendency of end-use electricity and hydrogen (long-term uncertainty). The uncertainty of renewable generation is all handled as predictive error, known as short-time uncertainty. However, the multi-year uncertainty of renewable generation plays a more critical role in long-run since ammonia production is almost always determined by renewable generation in such an isolated power system.

To the best of our knowledge, there is little or no work on the sizing of IRePtA systems when ammonia demand (i.e., actual annual output of ammonia) can also be optimized. Furthermore, considering the multi-year uncertainty of renewable generation, long-run risk is still underexplored.

To fill the above-mentioned gaps, this paper proposes an IGDT-MILFP model for IRePtA system planning. First, LCOA is directly formulated as the objective, which is a mixed integer linear fractional programming (MILFP) problem, and a corresponding solution approach is proposed. Then, information gap decision theory (IGDT) is utilized to model the uncertainty of renewable generation. Risk-averse and risk-seeking strategies are used to obtain robust and opportunistic planning results. Finally, Markov Chain Monte Carlo (MCMC) sampling-based posteriori analysis is presented to quantify the long-run performance. 

\subsection{Contributions}
\label{sec:contributions}
Specifically, the following contributions are made in this paper:

1) {\bf{Modeling Level}}: The MILFP model is established for IRePtA system planning for the first time. We reveal that the MILFP model is necessary and accurate to trade off investment and return compared with the existing MILP model-based approach when both source allocation and load demand are to be optimized. Furthermore, the IDGT model is adopted for handling the multi-year uncertainty of renewable generation, to avoid the long-run risk of IRePtA. 

2) {\bf{Solution Approach Level}}: A combined Charnes-Cooper (C\&C) transformation and Branch-and-Bound (B\&B) method is proposed to efficiently solve such a large-scale MILFP model. Compared to directly using a professional MINLP solver, such as \emph{gurobi}, the proposed solution approach can reduce the computational burden by orders of magnitude. 

3) {\bf{Application Level}}: It is revealed that exploiting the flexibility of ammonia synthesis is essential to reduce the LCOA of IRePtA systems. Using real data from Inner Mongolia, the planning results indicate that a nearly $43.53\%$ LCOA reduction is realized when ammonia flexibility is improved from a yearly to a daily regulation level with a competitive LCOA of 3610 RMB/t, even compared to coal-based ammonia. 



\section{Problem Description}
\label{sec:description}
In this section, IRePtA systems are introduced with the corresponding configuration and operation mode. Then, the necessity of using the MILFP model is discussed.

\subsection{Characteristics and Configuration of IRePtA Systems}
\label{sec:system_components}



\emph{1) Multi-timescale characteristics:} From the perspective of flexible response ability, BESS can quickly respond in several seconds \cite{zhao2020stochastic}, while the responsiveness of AE and AS are several minutes and hours \cite{2022arXiv221204754Y}, respectively. On the other hand, from the perspective of energy storage duration, the electricity storage of BESS is usually several hours, such as 2 h and 4 h \cite{zhao2020stochastic}. The hydrogen storage of the buffer tank is dozens of hours, while that of ammonia storage exceeds several weeks \cite{2022arXiv221204754Y}. Thus, multi-timescale issues of electricity, hydrogen, and ammonia are prominent, and an 8760 h time-sequence-operation-model is utilized to address them.

\emph{2) Minimum Power Supply for System Safety:} Both alkaline electrolysis and ammonia synthesis have minimum load constraints. Some auxiliary devices, such as compressors\cite{hu2022comprehensive} and circulating pumps \cite{olivier2017low}, are required to operate continuously for safety reasons. On the other hand, the emergency power supply of IRePtA is characterized by low frequency and long duration due to the intermittence of renewable generation. To this end, using PEMFC is an appropriate choice since hydrogen is available with large-scale hydrogen production and storage. 

\emph{3) Multi-year Uncertainty of Renewable Generation:} From the analysis of real data shown in Fig. \ref{fig:Multiyear_RG}, there exists considerable fluctuations of renewable generation at the yearly level. Taking the 20-year average FLH as a base, FLH of wind power varied between $-7.26\%$ and $7.39\%$, and that of solar power is $-3.24\% \sim 2.93\%$ (the third year's value is eliminated as an extreme climate data). In IRePtA systems, the annual output of ammonia is totally determined by renewable generation, which causes the long-run risk of a relatively smooth ammonia supply to downstream markets for 20 years. Thus, the IGDT method is utilized to address such long-term uncertainty. 

Based on the above engineering characteristics, a generic configuration of the IRePtA system is proposed, shown in Fig. \ref{fig:Topu_IRePtA}. The power source includes wind and solar power. The load demand consists of two parts. One is hydrogen production using alkaline electrolysis (AE) technology. The other is ammonia production, including nitrogen produced by pressure swing adsorption (PSA) and ammonia synthesis with Haber Bosch synthesis (HBS). The energy storages are hydrogen storage using a buffer tank, battery energy storage systems (BESS), and proton exchange membrane fuel cell (PEMFC), known as multiple energy storages (MESs). 




\subsection{Why is LCOA Selected as the Objective Function?}
\label{sec:difference_flexibility}
\subsubsection{Limited Flexibility of Ammonia Synthesis}
\label{sec:limited_flexibility_AS}
Our previous research \cite{2022arXiv221204754Y} demonstrated that the traditional operation mode of AS, i.e., \emph{8000-hour-full-load} operation as shown in the purple curve in Fig. \ref{fig:Ammonia_Adjustment_Fig}, is not adapted to the fluctuation of renewable generation. Therefore, a planned quasi-steady-state condition scheduling model for AS is proposed, as shown by the blue curve in Fig. \ref{fig:Ammonia_Adjustment_Fig}. Except for the general limitation of variation range and ramping speed, a more critical parameter related to AS flexibility is the scheduling period $\Delta T_{\rm{AS}}$, which takes values above daily level to meet the requirements of relatively stable chemical production. 

\subsubsection{Trade-off between Load Flexibility and System Economy}
\label{sec:Tradeoff_AS}
Flexible regulation of AS causes a reduction in the actual annual output of ammonia, as the orange area in Fig. \ref{fig:Ammonia_Adjustment_Fig}, which decreases the economy of the IRePtA system. On the other hand, the more flexible the AS is, the fewer demands of energy storage are required and less curtailment of renewable generation is required, which improves the system's economy by decreasing the investment cost of renewable power systems. Hence, an optimal capacity utilization rate of AS is necessary to trade off the load's flexibility and the system's economy. 

In other words, the minimization of the total cost or the maximization of the net revenue described by the MILP/LP model, is not equivalent to the minimization of LCOA, since the actual annual output of ammonia will be optimized. Our work in Section \ref{sec:IGDT_MILFP_Model} focuses on addressing the issues by directly using LCOA as the objective function in the MILFP model to trade off the investment and return.




\section{IGDT-MILFP IRePtA Planning Model}
\label{sec:IGDT_MILFP_Model}
In this section, first, the deterministic-MILFP (D-MILFP) IRePtA planning model is established, including the planning and operation of the IRePtA system. Then, the IGDT model is formulated with risk-averse and risk-seeking strategies. Finally, the overall IGDT-MILFP is concluded with an analysis of solving difficulties.

\subsection{D-MILFP IRePtA Planning Model}
\label{sec:D_MILFP_IRePtA}

Given the planning horizon $N$ and step length $\Delta T$, the set of time intervals $\mathbb{T}$ is defined as $\mathbb{T} = \left\{ 0,1,\ldots,N-1 \right\}$. And the set of facilities of IRePtA systems $\Omega_{\rm{F}}$ is given as ${\Omega}_{\rm{F}} = \left\{ \rm{W},\rm{S},\rm{AE},\rm{FC},\rm{HS},\rm{AS},\rm{B}\right\}$. The mathematical formulation of the proposed D-MILFP IRePtA planning model can be written as follows.
\begin{equation}
   DLCOA \equiv \min LCOA \qquad \qquad \qquad \label{eq:D_OF} \\
\end{equation}
s.t.
\begin{subequations}
\begin{align}
   &Objective \ Fucntion \ Related: \nonumber\\
   &LCOA = \frac {\left(C_{\rm{inv}} +C_{\rm{O\&M}} + R_{\rm{oper}}\right)}{O_{\rm{A}}} \label{eq:D_OF1} \\
   &C_{\rm{inv}} = \sum_{j \in {\Omega}_{\rm{F}}} {\rm{CRF}}(r,Y_j) C_j I_j^{\rm{init}}\qquad \qquad \qquad \quad \label{eq:D_inv}\\
   &C_{\rm{O\&M}} = \sum_{j \in {\Omega}_{\rm{F}}} I_j^{\rm{O\&M}} C_j I_j^{\rm{init}}    \label{eq:D_OM} \\
   &R_{\rm{oper}} = \lambda_{\rm{deg}} \Delta T \sum_{t \in \mathbb{T}} P_{t}^{\rm{B,disc}}  \label{eq:D_Roper} \\
   &O_{\rm{A}} = c_{\rm{H2A}} \Delta T \sum_{t \in \mathbb{T}} q_{t}^{\rm{H_2,out}} \label{eq:D_OA}
\end{align}
\end{subequations}
\vspace{-12pt}
\begin{subequations}
\begin{align}
   &Constraints \ of \ Alkaline \ Electrolysis: \nonumber\\
   &P_{t}^{\rm{AE}} = \kappa_{\rm{AE}} q_{t}^{\rm{H_2,in}} \qquad \qquad \qquad \qquad \qquad \qquad  \quad\label{eq:D_AE_1}  \\
   &\underline{\eta}^{\rm{AE}} C_{\rm{AE}} \leq P_{t}^{\rm{AE}} \leq \overline{\eta}^{\rm{AE}} C_{\rm{AE}}, \forall t \in \mathbb{T} \label{eq:D_AE_2}
\end{align}
\end{subequations}
\vspace{-12pt}
\begin{subequations}
\begin{align}
   &Constraints \ of \ Fuel \ Cells: \nonumber\\
   &P_{t}^{\rm{FC}} = \kappa_{\rm{FC}} q_{t}^{\rm{H_2,out1}} \qquad  \qquad \qquad \qquad \qquad \qquad  \label{eq:D_FC_1} \\
   &\underline{\eta}^{\rm{FC}} C_{\rm{FC}} \leq P_{t}^{\rm{FC}} \leq \overline{\eta}^{\rm{FC}} C_{\rm{FC}}, \forall t \in \mathbb{T} \label{eq:D_FC_2}
\end{align}
\end{subequations}
\vspace{-12pt}
\begin{subequations}
\begin{align}
   &Constraints \ of \ Ammonia \ Synthesis: \nonumber\\
   &P_{t}^{\rm{AS}} = \kappa_{\rm{AS}} q_{t}^{\rm{H_2,out}} \label{eq:D_AS_1} \\
   &q_{\tau}^{\rm{H_2,out}} = q_{k}^{\rm{H_2,QSS}} + \left( q_{k}^{\rm{H_2,QSS}} - q_{k+1}^{\rm{H_2,QSS}} \right) e^{-\frac{\tau}{T_{\rm{trans}}}}, \nonumber \\
    &\forall k \in \mathbb{K}, \forall \tau \in \left[ k \Delta T_{\rm{AS}},(k+1) \Delta T_{\rm{AS}}\right) \label{eq:D_AS_2} \\
   &\underline{\eta}^{\rm{AS}} q_{}^{\rm{H_2,r}} \leq q_{t}^{\rm{H_2,out}} \leq \overline{\eta}^{\rm{AS}} q_{}^{\rm{H_2,r}} \label{eq:D_AS_3}\\
   &r_{-}^{\rm{AS}} q_{}^{\rm{H_2,r}} \leq q_{t+1}^{\rm{H_2,out}} -q_{t}^{\rm{H_2,out}}  \leq r_{+}^{\rm{AS}} q_{}^{\rm{H_2,r}}, \forall t \in \mathbb{T} \label{eq:D_AS_4}\\
   &q_{}^{\rm{H_2,r}} = \frac{C_{\rm{AS}}}{8000 c_{\rm{H2A}}} \label{eq:D_AS_5} \\
   &r_{\rm{AS}} = \frac{O_{\rm{A}}}{C_{\rm{AS}}} \leq 1 \label{eq:D_AS_6}
\end{align}
\end{subequations}
\vspace{-12pt}
\begin{subequations}
\begin{align}
   &Constraints \ of \ Hydrogen \ Storage: \nonumber\\
   &n_{t+1}^{\rm{HS}} = n_{t}^{\rm{HS}} + \left( q_{t}^{\rm{H_2,in}} - q_{t}^{\rm{H_2,out}} - q_{t}^{\rm{H_2,out1}} \right) \Delta T  \label{eq:D_HS_1} \\
   &\underline{\eta}^{\rm{HS}} C_{\rm{HS}} \leq n_{t}^{\rm{HS}} \leq \overline{\eta}^{\rm{HS}} C_{\rm{HS}}, \forall t \in \mathbb{T} \label{eq:D_HS_2} \\
   &\left. n_{t}^{\rm{HS}} \right|_{t=0} = \left. n_{t}^{\rm{HS}} \right|_{t=N} = 50\% C_{\rm{HS}}\label{eq:D_HS_3}
\end{align}
\end{subequations}
\vspace{-12pt}
\begin{subequations}
\begin{align}
   &Constraints \ of \ BESS: \nonumber\\
   &ESOC_{t+1}^{\rm{B}} = (1-\xi_{\rm{B}})ESOC_{t}^{\rm{B}} \nonumber\\
   & +  \left(\eta_{\rm{B}}P_{t}^{\rm{B,ch}} - \frac{1}{\eta_{\rm{B}}} P_{t}^{\rm{B,disc}} \right)\Delta T \label{eq:D_BES_1} \\
   &\underline{\eta}^{\rm{B}} C_{\rm{B}} \leq ESOC_{t}^{\rm{B}} \leq \overline{\eta}^{\rm{B}} C_{\rm{B}} \label{eq:D_BES_2} \\
   &\left. ESOC_{t}^{\rm{B}} \right|_{t=0} = \left. ESOC_{t}^{\rm{B}} \right|_{t=N} = 50\% C_{\rm{B}} \label{eq:D_BES_3} \\
   &0 \leq P_{t}^{\rm{B,ch}},P_{t}^{\rm{B,disc}} \leq \frac{C_{\rm{B}}}{H_{\rm{B}}}   \label{eq:D_BES_4} \\
   &0 \leq P_{t}^{\rm{B,ch}} \leq {M_{\rm{B}}} \delta_{t}^{\rm{B}}   \label{eq:D_BES_5} \\
   &0 \leq P_{t}^{\rm{B,disc}} \leq {M_{\rm{B}}}(1-\delta_{t}^{\rm{B}}) , \forall t \in \mathbb{T}  \qquad \qquad \qquad \label{eq:D_BES_6}
\end{align}
\end{subequations}
\vspace{-12pt}
\begin{subequations}
\begin{align}
   &Constraints \ of \ System \ Integration: \nonumber\\
   &P_{t}^{j} = C_{j} P_{t}^{j,\rm{sta}} , j \in \left\{\rm{W},\rm{S} \right\}, \forall t \in \mathbb{T}  \qquad \qquad \; \label{eq:D_RG}\\
   &P_{t}^{\rm{W}} + P_{t}^{\rm{S}} + P_{t}^{\rm{B,disc}} + P_{t}^{\rm{FC}} \nonumber \\
   &= P_{t}^{\rm{AE}} +P_{t}^{\rm{AS}}  + P_{t}^{\rm{B,ch}} + P_{t}^{\rm{curt}} \qquad \qquad \;  \label{eq:D_Power_Balance} \\
   &P_{t}^{\rm{curt}} \geq 0, \forall t \in \mathbb{T} \label{eq:pcurt} \\
   &C_{j} = N_{j} C_{j}^{0}, N_{j} \in \mathbb{Z}, j \in \left\{\rm{W},\rm{S},\rm{AE} \right\}  \qquad \qquad \;  \label{eq:D_Capa_I}
\end{align}
\end{subequations}
\vspace{-12pt}
\begin{subequations}
\begin{align}
   &Decision \ Variables: \nonumber\\
   &C_{\rm{W}},  C_{\rm{S}}, C_{\rm{AE}},  C_{\rm{HS}}, C_{\rm{FC}}, C_{\rm{B}} 
    \label{eq:D_dv_1} \\
   &P_{t}^{\rm{W}}, P_{t}^{\rm{S}}, P_{t}^{\rm{AE}}, P_{t}^{\rm{AS}}, P_{t}^{\rm{FC}} , P_{t}^{\rm{B,disc}}, P_{t}^{\rm{B,ch}}, P_{t}^{\rm{curt}}, \nonumber \\
   &q_{t}^{\rm{H_2,in}}, q_{t}^{\rm{H_2,out}}, q_{t}^{\rm{H_2,out1}}, n_{t}^{\rm{HS}},ESOC_{t}^{\rm{B}},\forall t \in \mathbb{T}\quad \  \label{eq:D_dv_2} \\
   &N_{\rm{W}}, N_{\rm{S}}, N_{\rm{AE}}, \delta_{t}^{\rm{B}}, \forall t \in \mathbb{T}\label{eq:D_dv_3} 
\end{align}
\end{subequations}
\vspace{-12pt}

The objective function of D-MILFP IRePtA planning, represented in (\ref{eq:D_OF1}), is the levelized cost of ammonia (LCOA) \cite{li2021co}, denoted as a fractional function. The numerator of LCOA is composed of three parts: annualized capital investment cost $C_{\rm{inv}}$, annual operation and maintenance costs $C_{\rm{O\&M}}$, and annual variable operation cost $R_{\rm{oper}}$. $C_{\rm{inv}}$ represented in (\ref{eq:D_inv}), is determined by the initial investment cost using capital recovery factor (CRF) \cite{mehrtash2020enhanced} related to the interest rate $r = 8\%$ and the facilities lifetime $Y_j$, which is determined by ${\rm{CRF}}(r,Y_j) = \frac{r(1+r)^{Y_j}}{(1+r)^{Y_j}-1}$. $C_{\rm{O\&M}}$ represented in (\ref{eq:D_OM}), is assumed to be a fixed proportion ($I_{j}^{\rm{O\&M}}$) of the initial investment \cite{li2021co}. $R_{\rm{oper}}$ represented in (\ref{eq:D_Roper}), is known as a penalty term of BESS degradation, to trade-off fast regulation of BESS and its cycle degradation cost \cite{zhao2020stochastic}. The denominator of LCOA is the actual annual output of ammonia $O_{\rm{A}}$, which is an optimized variable represented in (\ref{eq:D_OA}).

The operation model of AE is denoted in (\ref{eq:D_AE_1})-(\ref{eq:D_AE_2}), including energy conversion from power $P_{t}^{\rm{AE}}$ to hydrogen $q_{t}^{\rm{H_2,in}}$, and variation range limitation of $P_{t}^{\rm{AE}}$. The PEMFC converts hydrogen $q_{t}^{\rm{H_2,out1}}$ to power $P_{t}^{\rm{FC}}$ with variation range limitation in (\ref{eq:D_FC_1})-(\ref{eq:D_FC_2}). 
Based on our previous work \cite{2022arXiv221204754Y}, the limited flexibility model of ammonia synthesis is represented in (\ref{eq:D_AS_1})-(\ref{eq:D_AS_6}), including energy consumption model from ammonia $q_{t}^{\rm{H_2,out}}$ to power $P_{t}^{\rm{AS}}$ in (\ref{eq:D_AS_1}), 1-order dynamic regulation model in (\ref{eq:D_AS_2}), variation range limitation in (\ref{eq:D_AS_3}), and ramping limitation in (\ref{eq:D_AS_4}), as shown in Fig. \ref{fig:Ammonia_Adjustment_Fig}. More details can be found in \cite{2022arXiv221204754Y}. Moreover, the rated work condition of AS $q_{}^{\rm{H_2.r}}$ is defined in (\ref{eq:D_AS_5}), following the operation principle of traditional ammonia synthesis facilities, i.e., \emph{8000-hour-full-load} operation mode. And the capacity utilization rate of AS $r_{\rm{AS}}$ is no more than 1, denoted in (\ref{eq:D_AS_6}).

The operation models of MESs, i.e., HS and BESS, are represented in (\ref{eq:D_HS_1})--(\ref{eq:D_HS_3}) and (\ref{eq:D_BES_1})--(\ref{eq:D_BES_6}) respectively. Constraints (\ref{eq:D_HS_1}) and (\ref{eq:D_BES_1}) are the state space equations of MESs, with the related state of charge (SOC) constraints, i.e., energy storage $ESOC_{t}^{\rm{B}}$ and hydrogen storage $n_{t}^{\rm{HS}}$ shown in (\ref{eq:D_HS_2}) and (\ref{eq:D_BES_2}). The initial and terminal SOC is assumed to be the same, denoted in (\ref{eq:D_HS_3}) and (\ref{eq:D_BES_3}). Moreover, discharge power $P_{t}^{\rm{B,disc}}$ and charge power $P_{t}^{\rm{B,ch}}$ are limited by the capacity of BESS, represented in (\ref{eq:D_BES_4}). In addition, discharge or charge states are exclusive, constrained by the binary variables $\delta_{t}^{\rm{B}}$ in (\ref{eq:D_BES_5})--(\ref{eq:D_BES_6}).

Constraints related to system integration are established in (\ref{eq:D_RG})-(\ref{eq:D_Capa_I}). According to constraint (\ref{eq:D_RG}), the maximum output of wind and solar power is determined by the capacities of WTs and PVs, using the standardized historical data $P_{t}^{\rm{W,sta}}$ and $P_{t}^{\rm{S,sta}}$ as a reference. Energy balancing constraint (\ref{eq:D_Power_Balance}) presents the power flow among the source, load, and energy storage in the IRePtA system, with a non-negative limitation of the curtailment of renewable power $P_{t}^{\rm{curt}}$. According to constraint (\ref{eq:D_Capa_I}), capacities of WTs, PVs, and electrolyzers are limited by the standard sizes of a single wind turbine, the pad-mounted transformer, and a single electrolyzer, based on practical engineering experience. Therefore, integer variables $N_j$ are introduced to the model.

Finally, decision variables are listed in (\ref{eq:D_dv_1})--(\ref{eq:D_dv_3}), including sizing related variables in (\ref{eq:D_dv_1}), operation continuous variables in (\ref{eq:D_dv_2}), and discrete variables in (\ref{eq:D_dv_3}).


\subsection{IGDT-based IRePtA Planning Model}
\label{sec:IGDT_Model}

To handle the uncertainty of renewable generation, robust-IGDT and opportunistic-IGDT models \cite{7994723,daneshvar2020novel,7346507} are proposed. The uncertainty of wind and solar power are described as the robust region $\pmb{\mathcal{U}}$, determined by the uncertainty horizon ($\alpha$) and forecasted/estimated renewable power ($\hat{P}_{\rm{W}}^{(t)}$ and  $\hat{P}_{\rm{S}}^{(t)}$), as
\begin{align}
   &P_{\rm{W}}^{(t)} \in \pmb{\mathcal{U}}_{\rm{W}} \left( \alpha,\hat{P}_{\rm{W}}^{(t)} \right) =  \left\{ P_{\rm{W}}^{(t)} \left| {\left | {\frac{ P_{\rm{W}}^{(t)} - \hat{P}_{\rm{W}}^{(t)} }{\hat{P}_{\rm{W}}^{(t)}} } \right |} \leq \alpha \right. \right\} \label{eq:uncertainty_pw} \\
   &P_{\rm{S}}^{(t)} \in \pmb{\mathcal{U}}_{\rm{S}} \left( \alpha,\hat{P}_{\rm{S}}^{(t)} \right) =  \left\{ P_{\rm{S}}^{(t)} \left| {\left | {\frac{ P_{\rm{S}}^{(t)} - \hat{P}_{\rm{S}}^{(t)} }{\hat{P}_{\rm{S}}^{(t)}} } \right |} \leq \alpha \right. \right\} \label{eq:uncertainty_ps}
 \end{align}

\subsubsection{Robust-IGDT Model}
\label{sec:risk_averse}
For a risk-averse planning strategy, robust-IGDT model maximizes the uncertainty horizon $\alpha_r$ in (\ref{eq:IGDT_1}), while robustness LCOA ($RLCOA$) is bounded by means of a given deviation factor $\beta_r$ and $DLCOA$ in (\ref{eq:IGDT_2}), under the worst-case, i.e., uncertainty variables take their lower bounds as given in (\ref{eq:IGDT_3})-(\ref{eq:IGDT_4}).
\begin{equation}
    \max \; \alpha_{r}  \label{eq:IGDT_1}
\end{equation}
s.t.
\begin{subequations}
\begin{align}
   &RLCOA = LCOA \leq (1+\beta_{r})DLCOA  \label{eq:IGDT_2} \\
   &P^{\rm{W}}_{t} = (1-\alpha_{r})C_{\rm{W}}^{} P_{t}^{\rm{W},\rm{sta}} \label{eq:IGDT_3} \\
   &P^{\rm{S}}_{t} = (1-\alpha_{r})C_{\rm{S}}^{} P_{t}^{\rm{S},\rm{sta}} \label{eq:IGDT_4}  \\
   &\left( \ref{eq:D_OF1} \right) - \left( \ref{eq:D_Capa_I} \right) \label{eq:IGDT_5}
\end{align}
\end{subequations}

\subsubsection{Opportunistic-IGDT Model}
\label{sec:risk_seeking}
For a risk-seeking planning strategy, opportunistic-IGDT model is to minimize the uncertainty horizon $\alpha_o$ in (\ref{eq:IGDTO_1}), while opportunity LCOA ($OLCOA$) is bounded by means of a given deviation factor $\beta_o$ and $DLCOA$ in (\ref{eq:IGDTO_2}), under the best-case, i.e., uncertainty variables take their upper bounds as given in (\ref{eq:IGDTO_3})--(\ref{eq:IGDTO_4}).
\begin{equation}
    \min \; \alpha_{o}  \label{eq:IGDTO_1}
\end{equation}
s.t.
\begin{subequations}
\begin{align}
   &OLCOA = LCOA \leq (1-\beta_{o})DLCOA  \label{eq:IGDTO_2} \\
   &P^{\rm{W}}_{t} = (1+\alpha_{o})C_{\rm{W}}^{} P_{t}^{\rm{W},\rm{sta}} \label{eq:IGDTO_3} \\
   &P^{\rm{S}}_{t} = (1+\alpha_{o})C_{\rm{S}}^{} P_{t}^{\rm{S},\rm{sta}} \label{eq:IGDTO_4}  \\
   &\left( \ref{eq:D_OF1} \right) - \left( \ref{eq:D_Capa_I} \right) \label{eq:IGDTO_5}
\end{align}
\end{subequations}
\noindent
where $DLCOA$ is the optimal solution of D-MILFP model (\ref{eq:D_OF}), which is known as a risk-neutral planning strategy. Although $LCOA$ is a fractional function, its denominator $O_{\rm{A}}$ is strictly greater than 0, therefore constraints (\ref{eq:IGDT_2}) and (\ref{eq:IGDTO_2}) can be easily transformed to a linear constraint. 

\subsection{Overall Planning Model and Analysis}
\label{sec:Overall_Model}

In other words, D-MILFP model (\ref{eq:D_OF}) and IGDT model (\ref{eq:IGDT_1})\&(\ref{eq:IGDTO_1}) both constitute the proposed IGDT-MILFP model. The scales of subproblems are listed in Table \ref{tab:scale_overall_opt}, with hundreds of thousands of decision variables and constraints. The proposed model (\ref{eq:D_OF})-(\ref{eq:D_Capa_I}) is a typical large-scale mixed-integer nonlinear programming (MINLP) model, which is also known as MILFP problem, and the aim is to optimize the ratio of two linear functions in the presence of linear constraints. It is difficult to solve such a large-scale MILFP model directly. Therefore, a combined C\&C and B\&B method is proposed in Section \ref{sec:solving_MILFP}. Moreover, the IGDT models (\ref{eq:IGDT_1}) and (\ref{eq:IGDTO_1}) are both MINLP problems with bilinear terms, which are handled in Section \ref{sec:solving_IGDT} by using a big M-based reformulation-linearization method.

\begin{table}[t]\scriptsize
  \renewcommand{\arraystretch}{2.0}
  \caption{The Scale of the Proposed Optimization Problem}
  \label{tab:scale_overall_opt}
  \centering
  \setlength{\tabcolsep}{3.5 pt}{
  \begin{tabular}{cccccc}
  \hline \hline
  ${}$   &\tabincell{c}{Programming  \\ Form}
  &\tabincell{c}{Continuous  \\ Variables}
  &\tabincell{c}{Discrete \\ Variables}
  &\tabincell{c}{Constraints}
  &\tabincell{c}{Bilinear \\ Terms}  \\
  \hline
  D-MILFP model  &MILFP  &$122,662$     &$8,763$    &$28,9104$   &None \\
  IGDT model     &MINLP  &$123,027$     &$8,763$    &$28,1996$   &$2$ \\
  \hline \hline
  \end{tabular}
  }
\end{table}

\section{Solution Approach to the Proposed IGDT-MILFP RePtA Planning Model}
\label{sec:Solution_Approach}
In this section, first, a combined C\&C and B\&B method for solving the D-MILFP model is presented. Second, a big M-based reformulation-linearization method is proposed to solve the IGDT model with bilinear terms. Then, an MCMC sampling-based posteriori analysis is proposed to quantify the long-run performance of the IGDT-MILFP model's solutions under different deviation factors. Finally, the framework of the proposed method is given.

\subsection{Solution Approach for D-MILFP Model: A Combined C\&C and B\&B Algorithm}
\label{sec:solving_MILFP}
Without loss of generality, the proposed D-MILFP model (\ref{eq:D_OF}) is rewritten as a general form of MILFP, denoted as the following problem \textbf{(P)}:
\begin{align}
   {\bf{(P)}}\quad \mathop {\min }\limits_{\bm{x},\bm{y}} \; & \frac{p_0+\sum\limits_{i \in \mathbb{I}}p_{1,i}x_i+\sum\limits_{j \in \mathbb{J}}p_{2,j}y_j}{q_0+\sum\limits_{i \in \mathbb{I}}q_{1,i}x_i+\sum\limits_{j \in \mathbb{J}}q_{2,j}y_j} \qquad \qquad \qquad \qquad \nonumber \\
   \rm{s.t.}&\sum\limits_{i \in \mathbb{I}}A_{1,i,m}x_i+\sum\limits_{j \in \mathbb{J}}A_{2,j,m}y_j \leq B_{m}, \forall m \in \mathbb{M} \nonumber\\
   &q_0 + \sum\limits_{i \in \mathbb{I}}q_{1,i}x_i + \sum\limits_{j \in \mathbb{J}}q_{2,j}y_j >0 \nonumber\\
   &x_i^L \leq x_i \leq x_i^U, x_i \in \mathbb{R}, \forall i \in \mathbb{I} \nonumber \\
   &y_j^L \leq y_j \leq y_j^U, y_j \in \mathbb{Z}, \forall j \in \mathbb{J} \label{eq:P_1}
\end{align}

\emph{Remark1:}  For an easy description, some equality constraints in (\ref{eq:D_OF})-(\ref{eq:D_Capa_I}) are represented as inequalities in (\ref{eq:P_1}), and binary variables are uniformly expressed as integer variables. Note that there is no loss of generality as one equality (e.g., $a = b$) can always be equivalently replaced by a pair of inequalities (e.g., $a \leq b$ and $a \geq b$), and for a binary variable $c$, constraints $0 \leq c \leq 1$ are added.



To address problem \textbf{(P)}, a combined C\&C and B\&B algorithm is proposed. Before introducing the algorithm, a lemma is given as follows.

\textbf{Lemma 1.} Linear relaxation of problem \textbf{(P)}, i.e., problem \textbf{(RP)}, known as the linear fractional programming (LFP), can be transformed to an equivalent linear programming (LP).

\textbf{Proof.} First, the linear relaxation of problem \textbf{(P)} is defined as problem \textbf{(RP)}, denoted as
\begin{align}
   {\bf{(RP)}} \quad \mathop {\min }\limits_{\bm{x},\bm{y}} \; & \frac{p_0+\sum\limits_{i \in \mathbb{I}}p_{1,i}x_i+\sum\limits_{j \in \mathbb{J}}p_{2,j}y_j}{q_0+\sum\limits_{i \in \mathbb{I}}q_{1,i}x_i+\sum\limits_{j \in \mathbb{J}}q_{2,j}y_j} \qquad \qquad \qquad \qquad \nonumber \\
   \rm{s.t.}&\sum\limits_{i \in \mathbb{I}}A_{1,i,m}x_i+\sum\limits_{j \in \mathbb{J}}A_{2,j,m}y_j \leq B_{m}, \forall m \in \mathbb{M} \nonumber\\
   &q_0 + \sum\limits_{i \in \mathbb{I}}q_{1,i}x_i + \sum\limits_{j \in \mathbb{J}}q_{2,j}y_j >0 \nonumber\\
   &x_i^L \leq x_i \leq x_i^U, x_i \in \mathbb{R}, \forall i \in \mathbb{I} \nonumber \\
   &y_j^L \leq y_j \leq y_j^U, y_j \in \mathbb{R}, \forall j \in \mathbb{J} \label{eq:P_2}
\end{align}
\noindent
Then, an auxiliary variable $u$ is introduced in (\ref{eq:CCT_1}), and then two variables $z_i$ and $w_j$ are introduced in (\ref{eq:CCT_2}) and (\ref{eq:CCT_3}) respectively, to sufficiently address the bilinear terms.
\begin{align}
   &u=\frac{1}{ q_0 + \sum\limits_{i \in \mathbb{I}}q_{1,i}x_i + \sum\limits_{j \in \mathbb{J}}q_{2,j}y_j } \label{eq:CCT_1} \\
   &z_i = u x_i, \forall i \in \mathbb{I} \label{eq:CCT_2} \\
   &w_j = u y_j, \forall j \in \mathbb{J} \label{eq:CCT_3}
\end{align}
\noindent
After the previous steps, problem \textbf{(RP)} can be transformed to LP form, as follows problem \textbf{(TP)}:
\begin{small}
\begin{align}
   {\bf{(TP)}}\mathop {\min }_{\bm{z},\bm{w},u} \; & {p_0 u + \sum\limits_{i \in \mathbb{I}}p_{1,i}z_i+\sum\limits_{j \in \mathbb{J}}p_{2,j}w_j} \qquad \qquad  \nonumber \\
   \rm{s.t.}&\sum\limits_{i \in \mathbb{I}}A_{1,i,m}z_i+\sum\limits_{j \in \mathbb{J}}A_{2,j,m}w_j - B_{m} u \leq 0, \forall m \in \mathbb{M} \nonumber\\
   &q_0 u + \sum\limits_{i \in \mathbb{I}}q_{1,i}z_i + \sum\limits_{j \in \mathbb{J}}q_{2,j}w_j = 1 \nonumber\\
   &u \geq 0, z_i \geq 0, w_j \geq 0, \forall i \in \mathbb{I}, \forall j \in \mathbb{J} \label{eq:P_3}
\end{align}
\end{small}%
\noindent 
Finally, (\ref{eq:P_2})--(\ref{eq:P_3}) conclude the proof. $\blacksquare$ 

\emph{Remark2:} The transformation procedure was first proposed by A. Charnes and  W.W. Cooper in \cite{charnes1962programming}; thus, it is also named the Charnes-Cooper (C\&C) transformation in \cite{gao2015fast}. Strict proof of optimality can be seen in Chapter 11.4, ``Linear Fractional Programming'' in \cite{bazaraa2013nonlinear}, which is outside the scope of this paper. 

The Branch-and-Bound (B\&B) algorithm is a typical solution approach for MINLP, while MILFP is a particular class of MINLP. The subproblem in MILFP for each node is LFP, which can be transformed to an equivalent LP problem using C\&C transformation, which is proven in Lemma 1. Therefore, a combined C\&C and B\&B algorithm is proposed, the C\&C transformation is embedded in a B\&B framework, and the entire procedure is presented in Algorithm \ref{alg:B_B}. 
\begin{algorithm}
\caption{A Combined Charnes-Cooper (C\&C) Transformation and Branch-and-Bound (B\&B) Algorithm}
\label{alg:B_B}
\LinesNumbered
\KwIn{Set $Incumbent := +\infty$, $\bm{L} := \varnothing$}
\KwOut{Optimal solution $\bm{x}^*$, $\bm{y}^*$}
Relax MILFP problem (\textbf{P}) in (\ref{eq:P_1}) to get the LFP problem (\textbf{RP}) in (\ref{eq:P_2}) \;
Transform LFP to an equivalnet LP problem (\textbf{TP}) in (\ref{eq:P_3}), by applying Charnes-Cooper transformation (\ref{eq:P_2})--(\ref{eq:CCT_3}), and add node 1 to $\bm{L}$, i.e., $\bm{L} \leftarrow \{\bm{TP}\}$\;
\While{$\bm{L} \neq \varnothing$}{
    Select node $n$ in $\bm{L}$, and remove it from $\bm{L}$, i.e., $\bm{L} = \bm{L}\backslash \{ {\rm{node}} \; n\}$ \;
    Solve the LP subproblem in node $n$\;
    \eIf{subproblem is infeasible}{
        Abort and fathom node $n$\;
    }{
        Denoted optimal solution as $(\tilde{\bm{z}}^{(n)}, \tilde{\bm{w}}^{(n)}, {\tilde{u}}^{(n)})$\, and corresponding objective value is $obj$\;
        \eIf{$\tilde{\bm{w}}^{(n)}/{\tilde{u}}^{(n)}$ is an integer solution}{
            \If{$obj < Incumbent$}{
                Update $Incumbent \leftarrow obj$\;
                Restore current optimal solution: $\bm{x}^* \leftarrow \tilde{\bm{z}}^{(n)}/{\tilde{u}}^{(n)}$, $\bm{y}^*\leftarrow \tilde{\bm{w}}^{(n)}/{\tilde{u}}^{(n)}$ \;
                Remove all the nodes in $\bm{L}$ with their objective values $> Incumbent$ \;
            }
        }{
            Select the variable $w_{j^*}$, whose subscript is:
            $j^* = {\rm{argmax}} \left \{ \left. { \left |\frac{\tilde{w}_j^{(n)}}{\tilde{u}^{(n)}} - \left [ \frac{\tilde{w}_j^{(n)}}{\tilde{u}^{(n)}} \right ] \right |} \right | j \in \mathbb{J}  \right \}$\;
            Create two nodes based on the current one as: \
            node $\vert \bm{L} \vert +1 = $ node $n$ $\cup \left\{ w_{j^*} \geq  \left \lceil \frac{\tilde{w}_{j^*}^{(n)}}{\tilde{u}^{(n)}} \right \rceil u \; \right\}$ \
            node $\vert \bm{L} \vert +2 = $ node $n$ $\cup \left\{ w_{j^*} \leq  \left \lfloor \frac{\tilde{w}_{j^*}^{(n)}}{\tilde{u}^{(n)}} \right \rfloor u \right\}$ \
            $\bm{L} \leftarrow \bm{L} \cup \left\{\rm{node} \; \vert \bm{L} \vert +1, \rm{node}\;\vert \bm{L} \vert +2\right\}$
        }
    }
}
\end{algorithm}

\subsection{A Reformulation-linearization Algorithm for IGDT Model}
\label{sec:solving_IGDT}
Based on the analysis in Section \ref{sec:Overall_Model}, there are four bilinear terms in IGDT model (\ref{eq:IGDT_1}) and (\ref{eq:IGDTO_1}), i.e., $\alpha_{r}C_{\rm{W}}$, $\alpha_{r}C_{\rm{S}}$, $\alpha_{o}C_{\rm{W}}$, and $\alpha_{o}C_{\rm{S}}$. Take bilinear term $\alpha_{r}C_{\rm{W}}$ as an example, the proposed big-M based reformulation-linearization algorithm is presented as follows.





First, binary variables  $\{b_l^{\rm{W}}, l = 0,1,\ldots,N_{b}^{\rm{W}}\}$ are introduced to replace the integer variable $N_{\rm{W}}$, therefore $\alpha_{r}C_{\rm{W}}$ is further denoted as
\begin{align}
   \alpha_{r}C_{\rm{W}} = C_{\rm{W}}^0 \sum \limits_{l=0}\limits^{N_b^{\rm{W}}}2^l \alpha_{r} b_l^{\rm{W}} = C_{\rm{W}}^0 \sum \limits_{l=0}\limits^{N_b^{\rm{W}}}2^l \delta_l^{\rm{W,r}}
   \label{eq:R_L_1}
\end{align}
\noindent
where $\delta_l^{\rm{W,r}}$ is a continuous variable with following constraints:
\begin{align}
   -M b_l^{\rm{W}} \leq &\delta_l^{\rm{W,r}} \leq M b_l^{\rm{W}} \label{eq:R_L_2}\\
   \alpha_{r}-M (1-b_l^{\rm{W}}) \leq &\delta_l^{\rm{W,r}} \leq \alpha_{r} + M (1-b_l^{\rm{W}}) \label{eq:R_L_3}
\end{align}
\noindent

Thus, (\ref{eq:R_L_1}) -- (\ref{eq:R_L_3}) concludes the reformulation-linearization algorithm. When bilinear terms are all processed, IGDT models (\ref{eq:IGDT_1}) and (\ref{eq:IGDTO_1}) are transformed to the MILP problem, which can be solved by \emph{Gurobi}.

\subsection{MCMC Sampling-based Posteriori Analysis}
\label{sec:MCMC}
\begin{figure}[t]
  \centering
  \includegraphics[width=3.46in]{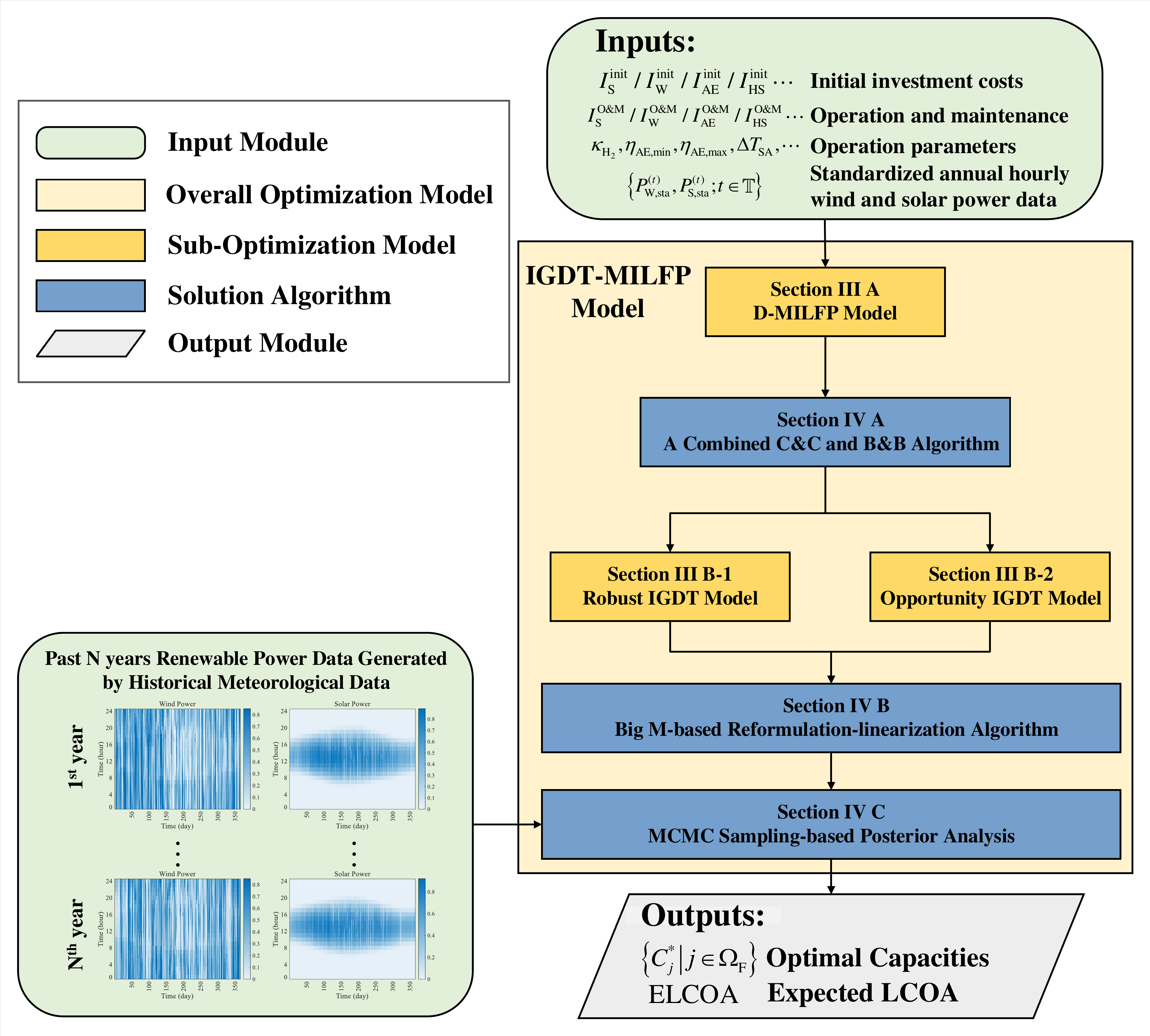}
  \caption{Framework of the solution approach for IGDT-MILFP model.}
  \label{fig:Algorithm_FlowChart}
\end{figure}

A posteriori analysis is proposed in this section to quantify the long-run performance of each IGDT-MILFP model's planning results, through a sufficiently large number of scenarios.  The full procedure is outlined below:

\emph{Step1)} Markov Chain Monte Carlo (MCMC) sampling method \cite{li2021markov} is used to generate posteriori renewable generation scenarios, based on the past 20 years' renewable power data generated by historical meteorological data, shown in the left bottom in Fig. \ref{fig:Algorithm_FlowChart}.

\emph{Step2)} For a specific planning result (i.e., capacities $C_j$ are given as the optimal solution $C_j^*$), each generated renewable generation scenario (i.e., $\left\{ \left ( P_{t,s}^{\rm{W}},P_{t,s}^{\rm{S}} \right ); t \in \mathbb{T} \right\}$) is used as input, the D-MILFP model is solved, and the corresponding $DLCOA_s$ is obtained. The above process is known as the stochastic production simulation of the IRePtA system.

\emph{Step3)} Since the stochastic production simulation result of the IRePtA system is obtained (i.e., $DLCOA_s, s=1,2,\ldots,N_{\rm{S}}$), the long-run performance index $ELOCA$ represents expected LOCA, denoted as
\begin{align}
   ELCOA = \frac{1}{N_{\rm{S}}} \sum \limits_{s=1}^{N_{\rm{S}}} DLCOA_s
   \label{eq:ELCOA}
\end{align}

The above procedure from step 1) to step 3) should be repeatedly run for all IGDT-MILFP model planning results under all given values of deviation factor $\beta_r / \beta_o$.

\subsection{Overall Solution Procedure and Application}
\label{sec:solving_framework}
The overall solution procedure is summarized as the flow chart, shown in Fig. \ref{fig:Algorithm_FlowChart}. 

\section{Case Studies}
\label{sec:case}
In this section, case studies are performed using the data of a real-life system in Inner Mongolia. First, the optimal planning results of the IGDT-MILFP model are presented with a detailed discussion. Second, the necessity and accuracy of the proposed MILFP model are demonstrated compared with the existing planning method by the MILP model. Then, the performances of different methods for solving the MILFP model are compared. Finally, a sensitivity analysis of ammonia flexibility is offered. 
\vspace{-12pt}
\subsection{Case Description and Setup}
\label{sec:descrip}

To study the proposed method, the IGDT-MILFP planning model in Section \ref{sec:IGDT_MILFP_Model} and corresponding solution approaches in Section \ref{sec:Solution_Approach}, are established in \emph{MATLAB R2020a} and solved by \emph{Gurobi 10.0.0}, environment on a desktop computer with Intel(R) Core (TM) i7-10700 CPU @ 2.90GHz processer.

The real-life system \cite{Batou-Project-2021} located in Baotou City, Inner Mongolia, China, shown in Fig. \ref{fig:Topu_IRePtA} is used in the case studies. The wind and solar power are generated by the real historical meteorology data from the project. And the 20-year average full load hours (FLH) of wind power is 3658 hours, and that of solar power is 1721 hours, as the yellow line shown in Fig. \ref{fig:Multiyear_RG}. The nominal annual output of ammonia is given as $C_{\rm{AS}}=10^5$ t. And, the investment, price, and operation parameters can be found in our previous work \cite{2022arXiv221204754Y}. And standard sizes of WT, PV, and AE are 6.25 MW, 3.15 MW, and 5 MW, respectively. Moreover, parameters related to BESS and PEMFC are listed as follows.

BESS: $I_{\rm{B}}^{\rm{init}} = 1,800 \rm{RMB/kWh}$, $I_{\rm{B}}^{\rm{O\&M}} = 1\%$, $Y_{\rm{B}} = 15$, $\xi_{\rm{B}} = 0.02\%$, $\eta_{\rm{B}} = 95\%$, $\underline{\eta}_{\rm{B}} = 10\%$, and $\overline{\eta}_{\rm{B}} = 90\%$ \cite{giraldo2018microgrids}.

PEMFC: $I_{\rm{FC}}^{\rm{init}} = 5,000 \rm{RMB/kW}$, $I_{\rm{FC}}^{\rm{O\&M}} = 2\%$, $Y_{\rm{FC}} = 15$, $\kappa_{\rm{FC}} = 1.5{\rm{kWh/Nm^3}}$, $\underline{\eta}_{\rm{FC}} = 0\%$, and $\overline{\eta}_{\rm{FC}} = 100\%$.
\vspace{-12pt}
\subsection{Optimal Planning Results of IGDT-MILFP Model}
\label{sec:IGDT_MILFP_results}
In this section, the optimal planning results of the IGDT-MILFP model are presented. Case 0 represents the result of the D-MILFP model, shown as risk-neutral planning. Cases 1--5 are the results of the robust-IGDT model, known as risk-averse planning. And cases 6--10 are the results of the opportunistic-IGDT model, with risk-seeking planning. All results above are summarized in Table \ref{tab:bs_bm_pm}.

\subsubsection{Optimal Deterministic Planning Results for $\beta = 0$}
\label{sec:Deterministic_results}
In case 0, the optimal planning result is obtained from the proposed D-MILFP model, with optimal objective value $DLCOA = 3630$ RMB/t. And corresponding $ELCOA = 3646$ RMB/t, which is found that $DLCOA < ELCOA$. It is worthwhile to note that $ELCOA$ simulates the behavior of the actual levelized cost of ammonia in long-run. The fact that ELCOA is greater than DLOCA indicates that deterministic planning cannot be averse to the risk in long-run, known as the multi-year uncertainty of renewable generation. Therefore, the IGDT model is essential to handle such uncertainty in long-run, as introduced in the following sections.

\begin{table*}[t]\scriptsize
  \renewcommand{\arraystretch}{2.0}
  \caption{Planning Results of IGDT-MILFP Model under Different Deviation Factors}
  \label{tab:bs_bm_pm}
  \centering
  \resizebox{1.9\columnwidth}{!}{
  \begin{threeparttable}[b]
  \begin{tabular}{cccccccccc}
  \hline \hline
  Case
  &$\beta_{r}$\tnote{b}\;\;/\;$\beta_{o}$\tnote{c}
  &\tabincell{c}{Optimal capacity\tnote{d}\\(MW, MW, MW, $\rm{Nm^3}$, MW, MWh)}
  &\tabincell{c}{Optimal \\ quantity\tnote{e}}
  &\tabincell{c}{$C_{\rm{inv,tot}}$\\(billion RMB)}
  &\tabincell{c}{$RLCOA$\tnote{b}\;/\ \\ $OLCOA$\tnote{c}\\(RMB/t)}
  &\tabincell{c}{$ELCOA$\\(RMB/t)}
  &\tabincell{c}{${\rm{FLH}}_{\rm{AE}}^*$}
  &\tabincell{c}{$r_{\rm{AS}^*}$}
  &\tabincell{c}{$\alpha_{r}$\tnote{b}\;\;/\;$\alpha_{o}$\tnote{c}} \\
  \hline

  case 0 &0\tnote{a}        &$\left\{ 218.75, 53.55,145,5.03 \times 10^5,8.80,14.84 \right\}$       &$\left\{ 35,17,29 \right\}$   &$2.271$         &$3630$\tnote{a}          &$3646$     &$5451$   &$79.99\%$   &$ 0$\tnote{a} \\
  case 1 &$0.02\tnote{b}$   &$\left\{ 225.00, 56.70,145,5.03 \times 10^5,8.79,15.07 \right\}$       &$\left\{ 36,18,29 \right\}$   &$2.315$         &$3702$\tnote{b}          &$3625$     &$5435$   &$79.75\%$   &$ 0.0378$\tnote{b} \\
  case 2 &$0.04\tnote{b}$   &$\left\{ 237.50, 47.25,145,5.19 \times 10^5,8.67,17.34 \right\}$       &$\left\{ 38,15,29 \right\}$   &$2.344$         &$3775$\tnote{b}          &$3635$     &$5397$   &$79.18\%$   &$ 0.0730$\tnote{b} \\
  case 3 &$0.06\tnote{b}$   &$\left\{ 256.25, 44.10,150,5.32 \times 10^5,8.92,17.45 \right\}$       &$\left\{ 41,14,30 \right\}$   &$2.446$         &$3847$\tnote{b}          &$3687$     &$5329$   &$80.88\%$   &$ 0.1038$\tnote{b} \\
  case 4 &$0.08\tnote{b}$   &$\left\{ 262.50, 44.10,150,5.45 \times 10^5,8.91,17.55 \right\}$       &$\left\{ 42,14,30 \right\}$   &$2.480$         &$3920$\tnote{b}          &$3703$     &$5295$   &$80.37\%$   &$ 0.1358$\tnote{b} \\
  case 5 &$0.10\tnote{b}$   &$\left\{ 268.75, 47.25,150,5.40 \times 10^5,8.92,17.40 \right\}$       &$\left\{ 43,15,30 \right\}$   &$2.523$         &$3993$\tnote{b}          &$3723$     &$5276$   &$80.08\%$   &$ 0.1651$\tnote{b} \\
  case 6 &$0.02\tnote{c}$   &$\left\{ 200.00, 69.30,140,4.72 \times 10^5,8.68,12.70 \right\}$       &$\left\{ 32,22,28 \right\}$   &$2.215$         &$3557$\tnote{c}          &$\bm{3610}$ &$5614$    &$79.53\%$   &$ 0.0427$\tnote{c} \\
  case 7 &$0.04\tnote{c}$   &$\left\{ 200.00, 56.70,145,4.85 \times 10^5,8.76,15.58 \right\}$       &$\left\{ 32,18,29 \right\}$   &$2.186$         &$3484$\tnote{c}          &$3617$     &$5495$    &$80.62\%$   &$ 0.0864$\tnote{c} \\
  case 8 &$0.06\tnote{c}$   &$\left\{ 187.50, 69.30,140,4.55 \times 10^5,8.64,13.37 \right\}$       &$\left\{ 30,22,28 \right\}$   &$2.148$         &$3412$\tnote{c}          &$3634$     &$5703$    &$80.79\%$   &$ 0.1346$\tnote{c} \\
  case 9 &$0.08\tnote{c}$   &$\left\{ 175.00, 81.90,140,4.40 \times 10^5,8.99,10.47 \right\}$       &$\left\{ 28,26,28 \right\}$   &$2.133$         &$3339$\tnote{c}          &$3697$     &$5782$    &$81.91\%$   &$ 0.1849$\tnote{c} \\
  case 10 &$0.10\tnote{c}$  &$\left\{ 168.75, 85.05,140,4.29 \times 10^5,9.18,9.74 \right\} $       &$\left\{ 27,27,28 \right\}$   &$2.111$         &$3267$\tnote{c}          &$3736$     &$5860$   &$83.02\%$   &$ 0.2385$\tnote{c} \\
  \hline \hline
  \end{tabular}

  \begin{tablenotes}
  \footnotesize
  \item[a,b,c] Related to risk-neutral, risk-averse and risk-seeking planning result,respectively.
  \item[d] Optimal size of WTs, PV, AE, HS, FC, and BES, respectively.
  \item[e] Optimal quantity of WTs, PV, and AE, respectively.
  \end{tablenotes}

  \end{threeparttable}
  }

\end{table*}

 \begin{figure}[t]
   \centering
   \includegraphics[width=3.46in]{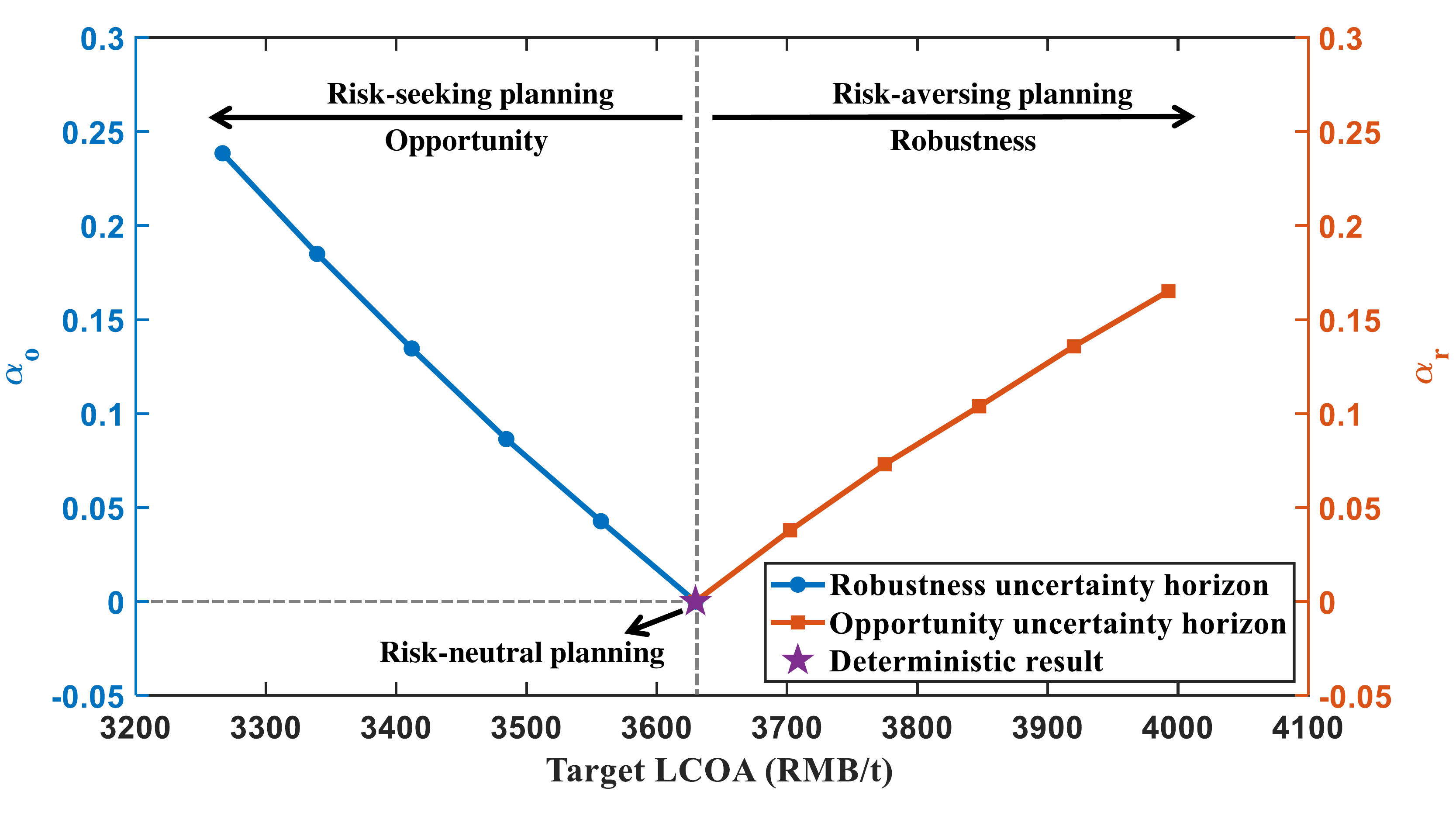}
   \caption{Robustness and opportunity uncertainty horizon for different LCOA targets.}
   \label{fig:alpha_TLCOA}
 \end{figure}
 
 \begin{figure}[t]
   \centering
   \includegraphics[width=3.46in]{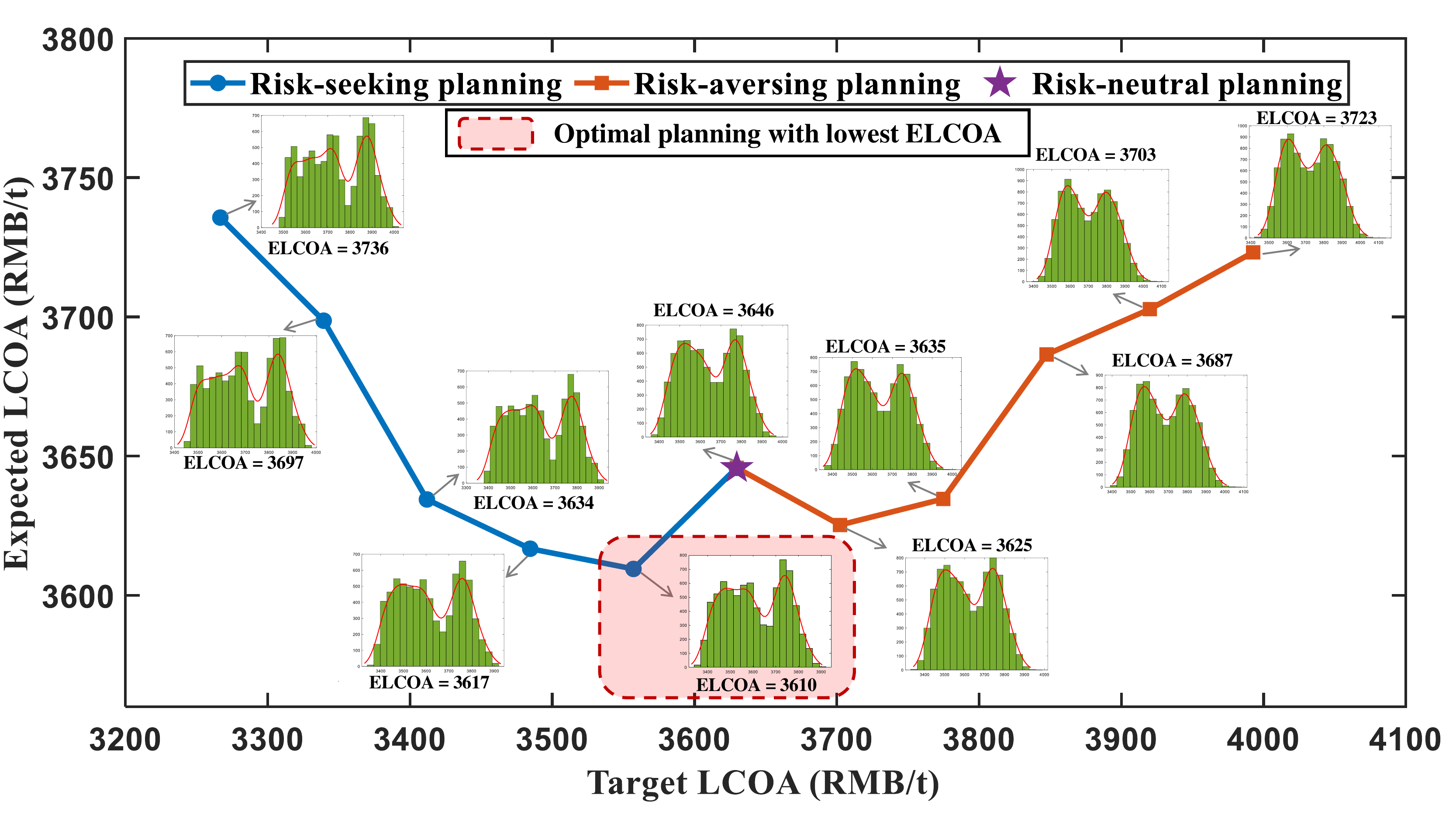}
   \caption{Expected LCOA with no information gap between the estimated and actual renewable generation against different LCOA targets.}
   \label{fig:ELCOA_TLCOA}
 \end{figure}

\subsubsection{Optimal Robust Planning Results for $\beta_r > 0$}
\label{sec:robust_results}
In cases 1--5, deviation factor $\beta_{r}$ is set as 0.02, 0.04, 0.06, 0.08, and 0.10, respectively. The corresponding maximum uncertainty horizon $\alpha_{r}$ are plotted as points in orange, shown in Fig. \ref{fig:alpha_TLCOA}, where the target LCOA is also known as RLCOA, calculated by $(1+\beta_{r})DLCOA$. Taking case 1 as example,  $\beta_{r}$ is 0.02, so the target LCOA (i.e., $RLCOA$) is $(1+0.02) \times 3630=3702$ RMB/t, and the maximum uncertainty horizon $\alpha_{r}$ is $0.0378$. This means that when the uncertainty of renewable generation is within $0.0378$, expected LCOA in long-run can be guaranteed to be no more than $3702$ RMB/t. The corresponding $ELCOA = 3625$ RMB/t is less than that of deterministic planning, which further indicates that the robust-IGDT model can be averse to risk in the long-run by reasonably increasing the capacities of WT and PV.

\subsubsection{Optimal Opportunistic Planning Results for $\beta_o > 0$}
\label{sec:opportunistic_results}
In cases 6--10, deviation factor $\beta_{o}$ is set as 0.02, 0.04, 0.06, 0.08, and 0.10, respectively. The corresponding minimum uncertainty horizon $\alpha_{o}$ is plotted as points in blue, as shown in Fig. \ref{fig:alpha_TLCOA}, where the target LCOA is also known as OLCOA, calculated by $(1-\beta_{o})DLCOA$. Take case 6 as an example, $\beta_{o}$ is 0.02, so target LCOA (i.e., $OLCOA$) is $(1-0.02) \times 3630=3557$ RMB/t, and minimum uncertainty horizon $\alpha_{r}$ is $0.0427$. It means that if $LCOA = 3557$ RMB/t is pursued, the minimum uncertainty horizon of renewable generation should exceed $0.0427$.  Moreover, the corresponding $ELCOA = 3610$ RMB/t is less than that of deterministic planning, indicating that the opportunistic-IGDT model can also reduce LCOA in the long run by reasonably decreasing the initial investment of facilities and seeking potential revenue in high-generation scenarios.

\subsubsection{Posteriori Analysis for Different Planning Results based on IGDT-MILFP Model}
\label{sec:posteriori_analysis_results}
To further compare cases 0--10, each case's $ELCOA$ obtained by the proposed posteriori analysis is plotted in Fig. \ref{fig:ELCOA_TLCOA}, with the corresponding distribution of $LCOA$ under 10,000 sampling renewable generation scenarios (i.e., $N_{\rm{S}} =10,000$), shown as histograms in green. 

By analyzing the distribution of $LCOA$ under different cases, we found that the robust-IGDT model can be averse to the risk of lower FLH of renewable generation, so the number of scenarios with $LCOA > DLCOA$ is reduced in case 1. And the opportunistic-IGDT model seeks the risk of higher FLH of renewable generation, so the number of scenarios with $LCOA < DLCOA$ increases in case 6. However, when the deviation factor is too large, the $ELCOA$ of the robust-model and the opportunistic-IGDT are both greater than that of the deterministic planning. It is because that much more robustness for planning may cause over-investment of facilities; for example, the initial investment cost of case 0 is $2.271$ billion RMB, while that of case 5 is $2.523$ billion RMB, which increased by $11.10\%$. In contrast, much more opportunity for planning may cause a lack of investment, leading to much more cost in low-generation scenarios.

In other words, from the target of pursuing the lowest expected LCOA in long-run, case 6 presents the best behavior with the lowest $ELCOA=3610$ RMB/t.

\vspace{-12pt}
\subsection{The Necessity and Accuracy of Proposed MILFP Model Compared with MILP Model}
\label{sec:MILFP_vs_MILP}
To demonstrate the necessity and accuracy of the proposed D-MILFP model, numerical simulations based on MILP are utilized, and the details are as follows.

\begin{figure}[t]
  \centering
  \includegraphics[width=3.46in]{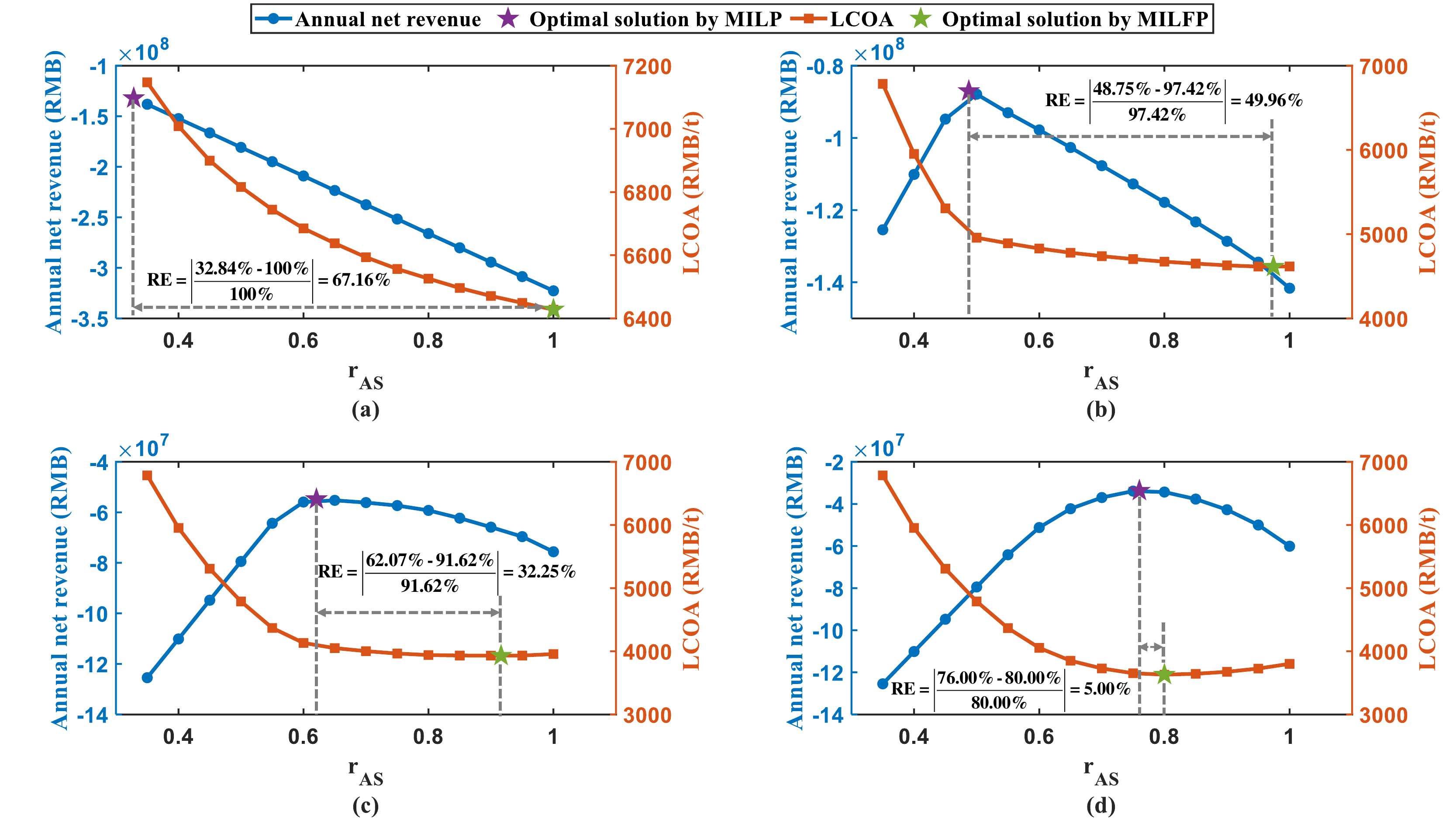}
  \caption{Optimal solution under different given $r_{\rm{AS}}$, compared with optimal solution of MILP model and proposed MILFP Model. (a) $\Delta T_{\rm{AS}}=1$ year; (b) $\Delta T_{\rm{AS}}=1$ month; (c) $\Delta T_{\rm{AS}}=1$ week; (d) $\Delta T_{\rm{AS}}=1$ day.}
  \label{fig:MILP_MILFP}
\end{figure}

The capacity utilization of ammonia synthesis $r_{\rm{AS}}$ is set from 0.35 to 1.00 with a step length of 0.05. For each given value of $r_{\rm{AS}}$, the D-MILFP model (\ref{eq:D_OF}) is transformed into an MILP model since the real annual output of ammonia $O_{\rm{A}}$ is determined to be constant based on the constraint (\ref{eq:D_AS_6}), i.e., $O_{\rm{AS}} = r_{\rm{AS}} C_{\rm{AS}}$. Two indices are calculated, first is annual net revenue $Reve$ defined in (\ref{eq:D_Reve}), shown on the left y-axis in Fig. \ref{fig:MILP_MILFP}. The corresponding ammonia price $\pi_{\rm{A}}$ is $3200 $ RMB/t \cite{CEIC-Economic-Database}. The other is LCOA defined in (\ref{eq:D_OF}), shown on the right y-axis in Fig. \ref{fig:MILP_MILFP}. 
\begin{align}
   &Reve = \pi_{\rm{A}} {O_{\rm{A}}} - \left(C_{\rm{inv}} +C_{\rm{O\&M}} + R_{\rm{oper}}\right)
   \label{eq:D_Reve} 
\end{align}

The above process is repeated at different flexibility levels of ammonia synthesis, i.e., yearly (fixed), monthly, weekly, and daily regulation levels. The corresponding results are plotted in Fig. \ref{fig:MILP_MILFP} (a)--(d). Furthermore, the green point shows the optimal solution with the lowest LCOA by the proposed MILFP model (\ref{eq:D_OF}). The purple point represents the optimal solution with the highest revenue by maximizing $Reve$ with the same constraint of the MILFP model, known as an MILP model.

From Fig. \ref{fig:MILP_MILFP}, the green and purple points are not coincident, indicating that maximizing annual net revenue by the MILP model is not equivalent to minimizing the LCOA by the proposed MILFP model. However, the latter result is the target of system planning. Furthermore, the relative error (RE) between the MILFP and MILP models is calculated. We found that the maximum RE is 67.16\% in Fig. \ref{fig:MILP_MILFP} (a), while the minimum RE is 5.00\% in Fig. \ref{fig:MILP_MILFP} (d). This means that less flexibility in ammonia synthesis leads to much more error between the MILFP and MILP models. Therefore the above results strongly prove the necessity and accuracy of the proposed MILFP model.

\vspace{-12pt}
\subsection{Performance Comparison of Different Methods for Solving MILFP Model}
\label{sec:Compared_CPUs}

\begin{table}[t]\scriptsize
   \renewcommand{\arraystretch}{2.0}
   \caption{Performance Comparison of Different Methods for Solving MILFP Model}
   \label{tab:computation_time}
   \centering
   \begin{threeparttable}[b]
   \begin{tabular}{ccccc}
   \hline \hline
   \multirow{2}{*}{Method}   &\multicolumn{2}{c}{Original Model} &\multicolumn{2}{c}{Modified Model} \\
   \cmidrule(r){2-3}  \cmidrule(r){4-5} \noalign{\smallskip}
   &\tabincell{c}{Objective\\ value} &\tabincell{c}{CPU \\ times (s)} &\tabincell{c}{Objective\\ value} &\tabincell{c}{CPU \\ times (s)} \\
   \hline
   \emph{Gurobi}  &\tabincell{c}{$4104.6188$\tnote{a} \\ (\;$29.3\% \tnote{b}$\;\;)}    &$18,000$    &\tabincell{c}{$3921.9760$\tnote{a} \\ (\;$33.9\% \tnote{b}$\;\;)}  &$18,000$ \\
   \tabincell{c}{Combined \\C\&C and B\&B}  &$3629.6732$     &$13,948$    &$\bm{3629.6732}$   &$\bm{385}$ \\
   \hline \hline
   \end{tabular}
   \begin{tablenotes}
   \footnotesize
   \item[a] Current suboptimal solution is obtained within the time limit (set as 18,000 seconds, i.e., 5 hours).
   \item[b] Current gap of suboptimal solution. 
   \end{tablenotes}
   \end{threeparttable}
 \end{table}


In this section, two cases are studied to demonstrate that BESS's discharging/charging state constraints can be relaxed since degradation of BESS is considered in the objective. One is the original MILFP model proposed in Section \ref{sec:D_MILFP_IRePtA}, and the other is a modified MILFP model without constraints (\ref{eq:D_BES_5})--(\ref{eq:D_BES_6}). Furthermore, to compare the performance of different algorithms for solving the MILFP model, two indices are introduced, i.e., objective value and computational CPU times, representing the accuracy and efficiency, respectively. And \emph{Gurobi} is used as a benchmarking method since \emph{Gurobi} claims that MINLP can be solved directly \cite{gurobi}.

The computational results corresponding to different models with different solution methods are summarized in Table \ref{tab:computation_time} and discussed in detail from the following two aspects.

1) Comparing the objective value of the original and modified models, when the proposed method, i.e., improved B\&B algorithm, is utilized, the optimal solution of the two models is obtained, with the same value of $3629.6732$. This indicates that discharge and charge state constraints (\ref{eq:D_BES_5})--(\ref{eq:D_BES_6}) can be relaxed without loss of optimality. Therefore, a large number of binary variables representing the discharging/charging state are reduced, significantly improving the solution efficiency. 

2) Comparing objective value and computational CPU times together. For the modified model, the combined C\&C and B\&B method can obtain the optimal solution in only 385 s. However, \emph{Gurobi} can only provide a suboptimal solution with a gap of $33.9\%$ within the time limit of 18,000 s. In conclusion, the proposed method presents a high efficiency to obtain the optimal solution, with nearly two orders of magnitude, reducing the computational burden.

\vspace{-12pt}
\subsection{Sensitivity Analysis of Ammonia Flexibility}
\label{sec:sensitivity_analysis}
\begin{figure}[t]
   \centering
   \includegraphics[width=3.4in]{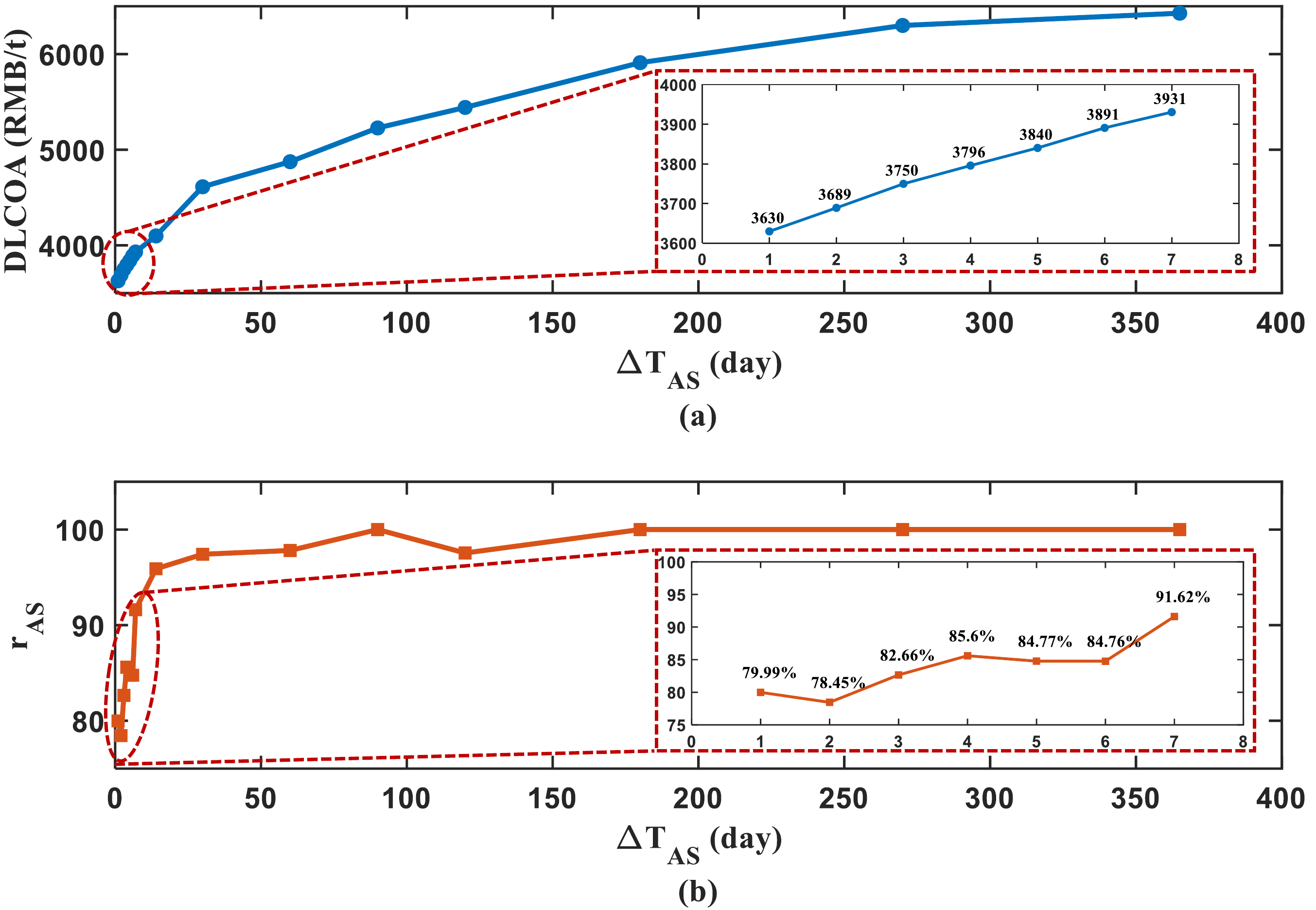}
   \caption{Sensitivity analysis of ammonia flexibility. (a) Optimal DLCOA. (b) Optimal capacity utilization of ammonia synthesis $r_{\rm{AS}}$.}
   \label{fig:Sensitivity_Ammonia_Flexibility}
   \vspace{-12pt}
 \end{figure}

The flexibility of ammonia is set from no flexibility ($\Delta T_{\mathrm{AS}}=1 $year) to high flexibility ($\Delta T_{\mathrm{AS}}=1 $ day), optimal $DLCOA$ are shown in Fig. \ref{fig:Sensitivity_Ammonia_Flexibility} (a), and corresponding optimal $r_{\rm{AS}}$ are plotted in \ref{fig:Sensitivity_Ammonia_Flexibility} (b).

From Fig. \ref{fig:Sensitivity_Ammonia_Flexibility} (a), with the improvement in the flexibility of ammonia synthesis, $DLCOA$ decreases. When the flexibility of ammonia synthesis is set as yearly, seasonally, monthly, weekly, and daily regulation levels, the corresponding LCOA are $6428$, $5228$, $4612$, $3931$, and $3630$ RMB/t, respectively. Therefore when ammonia synthesis is transformed from a constant load to a flexible load with daily regulation ability, LCOA will decrease from $6428$ to $3630$ RMB/t, decreasing by $43.53\%$. This means that improving the flexibility of ammonia synthesis plays a core role in reducing LCOA.

From Fig. \ref{fig:Sensitivity_Ammonia_Flexibility} (b), we found that when $\Delta T_{\mathrm{AS}}>1 $ month, i.e., the flexibility of ammonia is below the monthly level, the optimal $r_{\rm{AS}}$ is almost $100\%$ (varying between $97\%$ and $100\%$). However, with a further reduction of $\Delta T_{\mathrm{AS}}$, $r_{\rm{AS}}$ will significantly decrease, even reducing to $80\%$ when the daily regulation level is achieved. Furthermore, optimal $r_{\rm{AS}}$ under different flexibility levels of ammonia synthesis can be easily looked up in Fig. \ref{fig:Sensitivity_Ammonia_Flexibility} (b), which shows excellent value in engineering applications in IRePtA system planning.

\section{Conclusions}
\label{sec:conclusions}
An IGDT-MILFP model for IRePtA system planning is proposed in this paper, as well as the corresponding solution method. A real-life system in China's Inner Mongolia is studied. The results indicate that the MILFP model is necessary and accurate for IRePtA system planning. The proposed combined C\&C and B\&B method can reduce the computational burden by orders of magnitude for solving large-scale MILFP problems. Furthermore, the proposed method can well avoid long-run risk (i.e., multi-year uncertainty of renewable generation); therefore, the lowest expected LCOA is achieved.

Currently, the end product of the IRePtA system is only ammonia. However, when IRePtA systems participate in multiple markets like hydrogen and ammonia markets, the objective of the D-MILFP model should be well designed, and many more uncertainties, such as hydrogen and ammonia prices should be considered, which is one of the promising directions for future research.

\bibliographystyle{IEEEtran}
\bibliography{IEEEabrv,IRePtA_MILFP}

\begin{thebibliography}{10}
\providecommand{\url}[1]{#1}
\csname url@samestyle\endcsname
\providecommand{\newblock}{\relax}
\providecommand{\bibinfo}[2]{#2}
\providecommand{\BIBentrySTDinterwordspacing}{\spaceskip=0pt\relax}
\providecommand{\BIBentryALTinterwordstretchfactor}{4}
\providecommand{\BIBentryALTinterwordspacing}{\spaceskip=\fontdimen2\font plus
\BIBentryALTinterwordstretchfactor\fontdimen3\font minus
  \fontdimen4\font\relax}
\providecommand{\BIBforeignlanguage}[2]{{%
\expandafter\ifx\csname l@#1\endcsname\relax
\typeout{** WARNING: IEEEtran.bst: No hyphenation pattern has been}%
\typeout{** loaded for the language `#1'. Using the pattern for}%
\typeout{** the default language instead.}%
\else
\language=\csname l@#1\endcsname
\fi
#2}}
\providecommand{\BIBdecl}{\relax}
\BIBdecl

\bibitem{chehade2021progress}
G.~Chehade and I.~Dincer, ``Progress in green ammonia production as potential
  carbon-free fuel,'' \emph{Fuel}, vol. 299, p. 120845, 2021.

\bibitem{macfarlane2020roadmap}
D.~R. MacFarlane, P.~V. Cherepanov, J.~Choi, B.~H. Suryanto, R.~Y. Hodgetts,
  J.~M. Bakker, F.~M.~F. Vallana, and A.~N. Simonov, ``A roadmap to the ammonia
  economy,'' \emph{Joule}, vol.~4, no.~6, pp. 1186--1205, 2020.

\bibitem{20215}
\BIBentryALTinterwordspacing
``Danish partnership sets out to build world's first commercial scale green
  ammonia plant,'' \emph{Focus on Catalysts}, vol. 2021, no.~2, p.~5, 2021.
  [Online]. Available:
  \url{https://www.sciencedirect.com/science/article/pii/S1351418021000313}
\BIBentrySTDinterwordspacing

\bibitem{riera2022simulated}
J.~A. Riera, R.~M. Lima, I.~Hoteit, and O.~Knio, ``Simulated co-optimization of
  renewable energy and desalination systems in neom, saudi arabia,''
  \emph{Nature Communications}, vol.~13, no.~1, pp. 1--12, 2022.

\bibitem{salmon2021impact}
N.~Salmon and R.~Ba{\~n}ares-Alc{\'a}ntara, ``Impact of grid connectivity on
  cost and location of green ammonia production: Australia as a case study,''
  \emph{Energy \& Environmental Science}, vol.~14, no.~12, pp. 6655--6671,
  2021.

\bibitem{Batou-Project-2021}
T.~E.~B. of~Inner Mongolia Autonomous~Region, ``Notice of inner mongolia
  autonomous region energy bureau on carrying out the 2021 wind-solar hydrogen
  production integration demonstration,''
  \url{http://dbnyb.com/07/taiyangnen/2021/0827/51472.html}, 2021.

\bibitem{chiaramonti2019impacts}
D.~Chiaramonti and T.~Goumas, ``Impacts on industrial-scale market deployment
  of advanced biofuels and recycled carbon fuels from the eu renewable energy
  directive ii,'' \emph{Applied Energy}, vol. 251, p. 113351, 2019.

\bibitem{nayak2020techno}
R.~M. Nayak-Luke and R.~Ba{\~n}ares-Alc{\'a}ntara, ``Techno-economic viability
  of islanded green ammonia as a carbon-free energy vector and as a substitute
  for conventional production,'' \emph{Energy \& Environmental Science},
  vol.~13, no.~9, pp. 2957--2966, 2020.

\bibitem{rouwenhorst2019islanded}
K.~H. Rouwenhorst, A.~G. Van~der Ham, G.~Mul, and S.~R. Kersten, ``Islanded
  ammonia power systems: Technology review \& conceptual process design,''
  \emph{Renewable and Sustainable Energy Reviews}, vol. 114, p. 109339, 2019.

\bibitem{Batou-Project-2022}
{T. E. B. of Inner Mongolia Autonomous Region}, ``Notice of inner mongolia
  autonomous region energy bureau on carrying out the 2022 wind-solar hydrogen
  production integration demonstration,''
  \url{http://nyj.nmg.gov.cn/zwgk/zfxxgkzl/fdzdgknr/tzgg_16482/tz_16483/202209/t20220929_2143302.html},
  2022.

\bibitem{shepherd2022open}
J.~Shepherd, M.~H.~A. Khan, R.~Amal, R.~Daiyan, and I.~MacGill, ``Open-source
  project feasibility tools for supporting development of the green ammonia
  value chain,'' \emph{Energy Conversion and Management}, vol. 274, p. 116413,
  2022.

\bibitem{sanchez2018optimal}
A.~S{\'a}nchez and M.~Mart{\'\i}n, ``Optimal renewable production of ammonia
  from water and air,'' \emph{Journal of Cleaner Production}, vol. 178, pp.
  325--342, 2018.

\bibitem{klyapovskiy2021optimal}
S.~Klyapovskiy, Y.~Zheng, S.~You, and H.~W. Bindner, ``Optimal operation of the
  hydrogen-based energy management system with p2x demand response and ammonia
  plant,'' \emph{Applied Energy}, vol. 304, p. 117559, 2021.

\bibitem{armijo2020flexible}
J.~Armijo and C.~Philibert, ``Flexible production of green hydrogen and ammonia
  from variable solar and wind energy: Case study of chile and argentina,''
  \emph{International Journal of Hydrogen Energy}, vol.~45, no.~3, pp.
  1541--1558, 2020.

\bibitem{li2021co}
J.~Li, J.~Lin, P.-M. Heuser, H.~U. Heinrichs, J.~Xiao, F.~Liu, M.~Robinius,
  Y.~Song, and D.~Stolten, ``Co-planning of regional wind resources-based
  ammonia industry and the electric network: A case study of inner mongolia,''
  \emph{IEEE Transactions on Power Systems}, vol.~37, no.~1, pp. 65--80, 2021.

\bibitem{pan2021cost}
G.~Pan, W.~Gu, Q.~Hu, J.~Wang, F.~Teng, and G.~Strbac, ``Cost and low-carbon
  competitiveness of electrolytic hydrogen in china,'' \emph{Energy \&
  Environmental Science}, vol.~14, no.~9, pp. 4868--4881, 2021.

\bibitem{li2019optimal}
J.~Li, J.~Lin, H.~Zhang, Y.~Song, G.~Chen, L.~Ding, and D.~Liang, ``Optimal
  investment of electrolyzers and seasonal storages in hydrogen supply chains
  incorporated with renewable electric networks,'' \emph{IEEE Transactions on
  Sustainable Energy}, vol.~11, no.~3, pp. 1773--1784, 2019.

\bibitem{beerbuhl2015combined}
S.~S. Beerb{\"u}hl, M.~Fr{\"o}hling, and F.~Schultmann, ``Combined scheduling
  and capacity planning of electricity-based ammonia production to integrate
  renewable energies,'' \emph{European Journal of Operational Research}, vol.
  241, no.~3, pp. 851--862, 2015.

\bibitem{verleysen2020can}
K.~Verleysen, D.~Coppitters, A.~Parente, W.~De~Paepe, and F.~Contino, ``How can
  power-to-ammonia be robust? optimization of an ammonia synthesis plant
  powered by a wind turbine considering operational uncertainties,''
  \emph{Fuel}, vol. 266, p. 117049, 2020.

\bibitem{li2022coordinated}
J.~Li, J.~Lin, Y.~Song, J.~Xiao, F.~Liu, Y.~Zhao, and S.~Zhan, ``Coordinated
  planning of hvdcs and power-to-hydrogen supply chains for interregional
  renewable energy utilization,'' \emph{IEEE Transactions on Sustainable
  Energy}, 2022.

\bibitem{zhao2020stochastic}
X.~Zhao, X.~Shen, Q.~Guo, H.~Sun, and S.~S. Oren, ``A stochastic distribution
  system planning method considering regulation services and energy storage
  degradation,'' \emph{Applied Energy}, vol. 277, p. 115520, 2020.

\bibitem{2022arXiv221204754Y}
Z.~{Yu}, J.~{Lin}, F.~{Liu}, J.~{Li}, Y.~{Zhao}, Y.~{Song}, Y.~{Song}, and
  X.~{Zhang}, ``{Optimal Sizing and Pricing of Renewable Power to Ammonia
  Systems Considering the Limited Flexibility of Ammonia Synthesis},''
  \emph{arXiv e-prints}, p. arXiv:2212.04754, Dec. 2022.

\bibitem{hu2022comprehensive}
S.~Hu, B.~Guo, S.~Ding, F.~Yang, J.~Dang, B.~Liu, J.~Gu, J.~Ma, and M.~Ouyang,
  ``A comprehensive review of alkaline water electrolysis mathematical
  modeling,'' \emph{Applied Energy}, vol. 327, p. 120099, 2022.

\bibitem{olivier2017low}
P.~Olivier, C.~Bourasseau, and P.~B. Bouamama, ``Low-temperature electrolysis
  system modelling: A review,'' \emph{Renewable and Sustainable Energy
  Reviews}, vol.~78, pp. 280--300, 2017.

\bibitem{mehrtash2020enhanced}
M.~Mehrtash, F.~Capitanescu, P.~K. Heiselberg, T.~Gibon, and A.~Bertrand, ``An
  enhanced optimal pv and battery sizing model for zero energy buildings
  considering environmental impacts,'' \emph{IEEE Transactions on Industry
  Applications}, vol.~56, no.~6, pp. 6846--6856, 2020.

\bibitem{7994723}
M.~Ahmadigorji, N.~Amjady, and S.~Dehghan, ``A robust model for multiyear
  distribution network reinforcement planning based on information-gap decision
  theory,'' \emph{IEEE Transactions on Power Systems}, vol.~33, no.~2, pp.
  1339--1351, 2018.

\bibitem{daneshvar2020novel}
M.~Daneshvar, B.~Mohammadi-Ivatloo, K.~Zare, S.~Asadi, and A.~Anvari-Moghaddam,
  ``A novel operational model for interconnected microgrids participation in
  transactive energy market: A hybrid igdt/stochastic approach,'' \emph{IEEE
  Transactions on Industrial Informatics}, vol.~17, no.~6, pp. 4025--4035,
  2020.

\bibitem{7346507}
J.~Zhao, C.~Wan, Z.~Xu, and J.~Wang, ``Risk-based day-ahead scheduling of
  electric vehicle aggregator using information gap decision theory,''
  \emph{IEEE Transactions on Smart Grid}, vol.~8, no.~4, pp. 1609--1618, 2017.

\bibitem{charnes1962programming}
A.~Charnes and W.~W. Cooper, ``Programming with linear fractional
  functionals,'' \emph{Naval Research logistics quarterly}, vol.~9, no. 3-4,
  pp. 181--186, 1962.

\bibitem{gao2015fast}
J.~Gao and F.~You, ``Fast optimization algorithms for large-scale mixed-integer
  linear fractional programming problems,'' in \emph{2015 American Control
  Conference (ACC)}.\hskip 1em plus 0.5em minus 0.4em\relax IEEE, 2015, pp.
  5901--5906.

\bibitem{bazaraa2013nonlinear}
M.~S. Bazaraa, H.~D. Sherali, and C.~M. Shetty, \emph{Nonlinear programming:
  theory and algorithms}.\hskip 1em plus 0.5em minus 0.4em\relax John Wiley \&
  Sons, 2013.

\bibitem{li2021markov}
W.~Li, X.~Jia, X.~Li, Y.~Wang, and J.~Lee, ``A markov model for short term wind
  speed prediction by integrating the wind acceleration information,''
  \emph{Renewable Energy}, vol. 164, pp. 242--253, 2021.

\bibitem{giraldo2018microgrids}
J.~S. Giraldo, J.~A. Castrillon, J.~C. Lopez, M.~J. Rider, and C.~A. Castro,
  ``Microgrids energy management using robust convex programming,'' \emph{IEEE
  Transactions on Smart Grid}, vol.~10, no.~4, pp. 4520--4530, 2018.

\bibitem{CEIC-Economic-Database}
T.~University, ``Ceic economic database,''
  \url{https://ecollection.lib.tsinghua.edu.cn/databasenav/entrance/detail?mmsid=991021498963903966},
  2022.

\bibitem{gurobi}
\BIBentryALTinterwordspacing
{Gurobi Optimization, LLC}, ``{Gurobi Optimizer Reference Manual},'' 2022.
  [Online]. Available: \url{https://www.gurobi.com}
\BIBentrySTDinterwordspacing

\end{thebibliography}

\end{document}